\documentclass[a4paper,11pt]{article}

\usepackage{aas_macros}

\usepackage{jcappub} % for details on the use of the package, please see the JINST-author-manual
\usepackage{lineno}
\usepackage{inconsolata}

\usepackage{amsmath}

\usepackage{siunitx}
\DeclareSIUnit\year{yr}

\usepackage{natbib}
\defcitealias{gnedin1997probing}{GH98}

\newcommand{\aver}[1]{\langle#1\rangle}

\author[1]{Ivan Ridkokasha,}
 \emailAdd{ridkokasha@lorentz.leidenuniv.nl}
\affiliation[1]{%
 Lorentz Institute for Theoretical Physics, Leiden University, PO Box 9506, NL-2300 RA Leiden, The Netherlands
}

\author[1]{Andres Aramburo-Garcia,}
\emailAdd{aramburo@lorentz.leidenuniv.nl}

\author[2]{Kyrylo Bondarenko,}
 \emailAdd{kyrylo.bondarenko@su.se}
\affiliation[2]{
Nordita, KTH Royal Institute of Technology and Stockholm University, Hannes Alfv\'ens v\"ag
12, 10691 Stockholm, Sweden}

\author[3]{Anastasia Sokolenko,}
\emailAdd{sokolenko@wustl.edu}
\affiliation[3]{
Department of Physics and McDonnell Center for the Space Sciences, Washington University, St. Louis, Missouri 63130, USA
}

\author[1,4]{Matthieu Schaller,}
\emailAdd{mschaller@lorentz.leidenuniv.nl}
\affiliation[4]{Leiden Observatory, Leiden University, PO Box 9513, NL-2300 RA Leiden, The Netherlands}

\author[1]{Alexey Boyarsky}
 \emailAdd{boyarsky@lorentz.leidenuniv.nl}

\title{\boldmath 3D matter power spectrum correspondence to 1D Lyman-alpha flux power spectrum}
\abstract{
The 3D distribution of matter at small scales encodes valuable information about the nature of dark matter and other fundamental physics. A prominent probe of such scales outside galaxies is the Lyman-alpha forest, which studies absorption features in the spectra of high-redshift quasars caused by neutral hydrogen. The measured quantity is the power spectrum of the absorbed flux, which indirectly traces the underlying matter distribution.
However, the connection between the measured flux power spectrum and the underlying 3D dark matter power spectrum is highly nontrivial. The flux power spectrum (i) represents a one-dimensional projection of the density field; (ii) traces only neutral hydrogen, subject to thermodynamic pressure; and (iii) is a nonlinear function of local matter density. Additionally, thermal broadening and redshift-space distortions—determined not only by the hydrogen distribution but also by its thermal state and local velocity field—further complicate interpretation.
To robustly constrain dark matter properties using the Lyman-alpha forest, these systematics must be carefully modeled and controlled.
In this paper, we present a simple phenomenological recipe for mapping the 3D matter power spectrum to the flux power spectrum. We first motivate our approach in the linear regime, then extend it to later times and into the nonlinear regime.
We validate our model against a broad suite of warm and cold dark matter simulations, demonstrating that our recipe yields consistent and accurate estimates across a wide parameter space.
}

\begin{document}
\maketitle
%%%%%%%%%%%%%%%%%%%%%%%%%%%%%%%%%%%%%%%%%%%%%%%%%%
%%%%%%%%%%%%%%%%%%%%%%%%%%%%%%%%%%%%%%%%%%%%%%%%%%
\flushbottom
%%%%%%%%%%%%%%%%%%%%%%%%%%%%%%%%%%%%%%%%%%%%%%%%%%
%%%%%%%%%%%%%%%%%%%%%%%%%%%%%%%%%%%%%%%%%%%%%%%%%%
\section{Introduction}\label{sec:intro}
%%%%%%%%%%%%%%%%%%%%%%%%%%%%%%%%%%%%%%%%%%%%%%%%%%
%%%%%%%%%%%%%%%%%%%%%%%%%%%%%%%%%%%%%%%%%%%%%%%%%%

% \kb{maybe add a few words about fuzziness?}

The nature of dark matter (DM) remains one of the most pressing open questions in modern physics. With a density approximately five times greater than that of baryonic matter, DM plays a central role in the formation of cosmic structures~\cite{Bertone:2016nfn,Arbey:2021gdg}. Hence, understanding these structures can provide important clues that can help us constrain the specific properties of the DM. One of such properties could be the ``warmness'' of DM~\cite{Bode:2000gq}, meaning how and for how long the hypothetical DM particle was relativistic to later cool to non-relativistic velocities. During a relativistic period, the DM particles experience free streaming, which suppresses the formation of structures in the universe on small scales; the size of these structures is directly related to the warmness of DM. Therefore, observing and accurately measuring the size and distribution of cosmic structures is an important means to constrain dark matter warmness. 

The Lyman-alpha forest provides a unique window into the state of the intergalactic medium on the smallest scales at different redshifts, making it an ideal tool for studying the nature of DM~\cite{Hui:1996fh,Gnedin:2001wg,Boyarsky:2008xj}. In particular, the Lyman-alpha flux power spectrum (FPS) exhibits a cut-off at high wavenumbers (small scales) in the presence of warm dark matter (WDM), reflecting the suppression of small-scale structures due to free streaming~\cite{garzilli2021constrain}.

The challenge of such analysis is that other astrophysical effects can generate a similar cut-off in the Lyman-alpha FPS~\cite{1997MNRAS.292...27H,gnedin1997probing,Boyarsky:2008xj,2015MNRAS.450.1465G}. We do not observe dark matter directly but rather neutral hydrogen distribution, which reflects not only the distribution of DM but is also affected by the internal thermodynamic pressure of the baryons that prevent the collapse into small structures (pressure smoothing), which mimics the effect of WDM. Furthermore, we analyze the distribution of hydrogen by measuring absorption lines, which also exhibit thermal line broadening and redshift-space distortions that further impact the Lyman-alpha FPS. To connect theoretical models with observations, cosmological numerical simulations are the standard tool in the field~\cite{Villasenor:2022aiy,puchwein2023sherwood} as they self-consistently model the collapse of baryon and DM structures. However, to conduct a robust analysis of the Lyman-alpha flux power spectrum, it is advantageous to employ analytical tools capable of disentangling these complex effects.

Viel et al. (2005)~\cite{viel2005constraining} used Boltzmann solvers to parametrize the free-streaming effect of warm dark matter on the 3D matter power spectrum for a range of WDM masses. Their model predicts a scale-dependent cut-off in the power spectrum, which is solely determined by the mass of the WDM particle. Importantly, this cut-off is redshift-independent, as free streaming occurs only at early times, well before the epochs relevant to observations. See App.~\ref{app:wdm_effect} for more details.

The pressure smoothing effect, however, does not have such a simple redshift dependence. The simplest estimate of the pressure effect and its redshift dependence is given by the Jeans scale~\cite{Binney2008}. Solving linearized hydrodynamic equations in a static universe, one finds that for modes with smaller wavenumbers (larger scales), gravity pull is stronger than pressure, and they collapse, while for larger wavenumbers (smaller scales), pressure overcomes gravity, and modes do not collapse. 

The Jeans derivation neglects the time it takes for the pressure to actually expand the structure and, thus, overestimates the pressure effect. Gnedin \& Hui were the first to account for this~\cite{gnedin1997probing} (hereafter \citetalias{gnedin1997probing}), introducing the filtering scale $k_F$ in analogy to the Jeans scale. In the linear regime, they calculated the filtering scale as a function of the thermal history and redshift. In later works~\cite{gnedin2003linear}, using numerical methods, an approximation of the form of the cut-off in the 3D power spectrum was found. For more details see App.~\ref{app:pressure_effect}.

Kulkarni et al.~\cite{kulkarni2015characterizing} tested the \citetalias{gnedin1997probing} model at redshifts relevant to the Lyman-alpha observations, after nonlinear structures started to form. They found that the baryonic matter power spectrum does not follow the \citetalias{gnedin1997probing} prediction, in fact they found no cut-off in matter power spectrum. The discrepancy arises because small nonlinear structures add power to all scales\footnote{Similar to the Fourier transform of the Dirac delta function.} and erase any cut-off. It is thus, necessary to remove dense nonlinear structures from the analysis to restore the cut-off. As Kulkarni et al. showed, this is achieved automatically by Lyman-alpha observations. Because of the exponential relation between flux and optical depth, $F=e^{-\tau}$, dense structures saturate the optical depth and give zero flux, effectively removing themselves from the analysis. Therefore, at low redshifts, the Lyman-alpha FPS is better suited to measure small-scale cut-off in the matter PS than the direct measurements of the matter power spectrum itself. Using this insight, we build a semi-analytical theory of how to translate predictions in the linear regime for the 3D matter power spectrum to the 1D Lyman-alpha flux power spectrum and test it against a range of cosmological simulations.

This paper is organized as follows: in Sec.~\ref{sec:theory} we introduce our phenomenological ansatz for the prediction of the cut-off in the flux power spectrum. In Sec.~\ref{sec:simulations} we describe the cosmological simulations used in this work and the methods used for their analysis. Next, in Sec.~\ref{sec:results} we make a comparison between our theory and simulations. Finally, we conclude in Sec.~\ref{sec:conclusions}.
% \kb{put more clear goals and discuss the structure of the paper}
% To make future constraints on the warmness of dark matter more robust, we present in this work a semi-analytical model to predict WDM free streaming and pressure smoothing effects on the Lyman-alpha flux power spectrum. More precisely, we show how to convert calculations for the 3D matter power spectrum in linear regime to the 1D Lyman-alpha flux power spectrum in nonlinear mode.
%%%%%%%%%%%%%%%%%%%%%%%%%%%%%%%%%%%%%%%%%%%%%%%%%%
%%%%%%%%%%%%%%%%%%%%%%%%%%%%%%%%%%%%%%%%%%%%%%%%%%
\section{Theoretical Model}\label{sec:theory}
%%%%%%%%%%%%%%%%%%%%%%%%%%%%%%%%%%%%%%%%%%%%%%%%%%
%%%%%%%%%%%%%%%%%%%%%%%%%%%%%%%%%%%%%%%%%%%%%%%%%%

%%%%%%%%%%%%%%%%%%%%%%%%%%%%%%%%%%%%%%%%%%%%%%%%%%
%\subsection{Semi-analytical description of the 1D flux power spectrum}
%%%%%%%%%%%%%%%%%%%%%%%%%%%%%%%%%%%%%%%%%%%%%%%%%%
% Our approach consists in using the previous methods for describing the physical effects on the 3D power spectrum and connecting them with the 1D flux power spectrum. 
In this section, we introduce our model to predict the cut-off in the 1D flux power spectrum from pressure and warm dark matter effects. Consider a statistically isotropic and homogeneous three-dimensional field $f(\vec{r})$ with a dimensionless power spectrum given by $\Delta_\text{3D} = \frac{k^3 \aver{\tilde{f}_k^2}}{2\pi^2}$. If we observe it only along a single line of sight (e.g., along the $z$-axis), then the dimensionless power spectrum of the resulting one-dimensional field $F(z) = f(0,0,z)$ is~\cite{lumsden1989clustering}:

\begin{equation}
    \Delta_\text{1D}(k)=\frac{k\aver{\tilde F_k^2}}{\pi}=k\int^\infty_k \Delta_\text{3D}(q) \frac{dq}{q^2}.
    \label{eq:1D-3D-connection}
\end{equation}

The issue with this relationship is that even for small wavenumbers $k$, $\Delta_\text{1D}(k)$ inside the integral will be influenced by the largest $q$, i.e. the smallest scales in the 3D power spectrum. As a result, variations in dark-matter models in small-scale baryonic physics will impact the behavior of the 1D power spectrum across \emph{all} scales.

We address this by separating the integral into two different scales of $q$:
\begin{equation}\label{eq:1D3D_separated_scales}
    \Delta_\text{1D}(k)=k\int^{k_\mathrm{max}}_k \Delta_\text{3D}(q) \frac{dq}{q^2} + k\int^\infty_{k_\mathrm{max}} \Delta_\text{3D}(q) \frac{dq}{q^2}.
\end{equation}
The behavior of $\Delta_\text{3D}$ in the $k>k_\mathrm{max}$ regime is governed by complex small-structure baryonic physics and is sensitive to the particularities of the DM model. We will treat the whole integral over this region as an unknown constant $\xi$ and as a free parameter of our model. 

To model the 3D power spectrum at smaller wavenumbers $k\leq k_\mathrm{max}$ we will use previous works on WDM and pressure effects. Namely, Viel et al.~\cite{viel2005constraining} estimated the effect of WDM on the 3D matter power spectrum which we approximate in the following form (see App.~\ref{app:wdm_effect} for more details):

\begin{align}\label{eq:exponent_wdm}
    \frac{\Delta^\mathrm{3D}_\mathrm{WDM}}{\Delta^\mathrm{3D}_\mathrm{CDM}}&=\exp{\left(-2\frac{k^2}{k_\mathrm{WDM}^2}\right)},
\end{align}
where $k_\mathrm{WDM}$ is a corresponding cut-off scale that depends on the thermal relic particle mass and given by Eq.~\eqref{eq:kWDM}.

Similarly, we use the \citetalias{gnedin1997probing} estimate of the pressure effects in the linear regime and the form of the cut-off from~\cite{gnedin2003linear} (for more details see App.~\ref{app:pressure_effect}):
\begin{equation}\label{eq:exponent}
    \frac{\Delta^\mathrm{3D}_\mathrm{baryons}}{\Delta^\mathrm{3D}_\mathrm{DM}}=\exp{\left(-2\frac{k^2}{k_F^2}\right)},
\end{equation}
where the filtering scale $k_F$ is given by~\eqref{eq:kF}.

Multiplying formulas~\eqref{eq:exponent_wdm} and~\eqref{eq:exponent} we get the following form of the cut-off for the joint WDM and pressure effects:
\begin{align}
    \frac{\Delta^\mathrm{3D}_\mathrm{baryons}}{\Delta^\mathrm{3D}_\mathrm{CDM}}&=\exp{\left(-2\frac{k^2}{k_\mathrm{WDM}^2}\right)}\exp{\left(-2\frac{k^2}{k_F^2}\right)}=\exp{\left(-2\frac{k^2}{k_\mathrm{cut}^2}\right)},\nonumber\\
    \frac{1}{k_\mathrm{cut}^2}&=\frac{1}{k_\mathrm{WDM}^2}+ \frac{1}{k_F^2}.\label{eq:adding_effects}
\end{align}
This formula gives a simple method of combining the two effects, which we will test in Sec.~\ref{sec:results}. 

Finally, to model the cold dark matter power spectrum, we take a simplistic power law approximation $\Delta^\mathrm{3D}_\mathrm{CDM}=Ak^a$.\footnote{We will later see that this simple assumption works reasonably well for all our simulations.} Since in all cases the form of the cut-off is the same, we will use the following simple formula for pure WDM effect, pure pressure effect, or a combination of both:
\begin{equation}\label{eq:full_exponent}
   \Delta^\mathrm{3D}_\mathrm{baryons}=Ak^a\exp{\left(-2\frac{k^2}{k_\mathrm{cut}^2}\right)},
\end{equation}
where $k_\mathrm{cut}$ combines the different physical effects.

Now that we have an ansatz for the 3D matter power spectrum, we can substitute it back into Eq.~\eqref{eq:1D3D_separated_scales}. We also extend the integral to the infinity and correspondingly redefine the constant $\xi$.\footnote{We define $\xi$ to be the difference between the actual $\Delta_\mathrm{3D}$ and our model with exponential cut-off $\xi=\int^\infty_{k_\mathrm{max}} \left(\Delta_\text{3D}(q) - Aq^a\exp{\left(-2\frac{q^2}{k_\mathrm{cut}^2}\right)} \right) \frac{dq}{q^2}$.} With this, our model takes the following form:

\begin{equation}\label{eq:full_fit}
\boxed{
    \Delta_\text{1D}(k)=k\int^{\infty}_k Aq^a\exp{\left(-2\frac{q^2}{k_\mathrm{cut}^2}\right)} \frac{dq}{q^2} + k\cdot \xi,\ k\leq k_\mathrm{max}.}
\end{equation}

In our analysis, we will focus particularly on the cut-off scale term $k_{\text{cut}}$ since it encloses the pressure smoothing and the WDM free streaming effects. We will test whether these simple predictions, based on linear theory, would work for the nonlinear regime probed by simulations.

%%%%%%%%%%%%%%%%%%%%%%%%%%%%%%%%%%%%%%%%%%%%%%%%%%
%%%%%%%%%%%%%%%%%%%%%%%%%%%%%%%%%%%%%%%%%%%%%%%%%%
\section{Description of the simulations}\label{sec:simulations}
%%%%%%%%%%%%%%%%%%%%%%%%%%%%%%%%%%%%%%%%%%%%%%%%%%
%%%%%%%%%%%%%%%%%%%%%%%%%%%%%%%%%%%%%%%%%%%%%%%%%%
In order to test our theoretical model, we used an ensemble of smoothed particle hydrodynamics (SPH) hydrodynamic simulations. These simulations have different physical parameters that have a direct effect on the power spectrum cut-off: thermal history, DM model, and resolution. Among the simulations analyzed in this work, we included those described by Garzilli et al.~\cite{garzilli2019lyman,garzilli2021constrain} and the rest were performed using \textsc{swift}~\cite{SWIFT}, a modular highly-parallel open-source code, that couples and solves hydrodynamics, gravity, cosmology and galaxy formation. 
For these, we adopted the cosmological parameters from the latest Dark Energy Survey paper~\cite{abbott2022dark}.\footnote{In Dark Energy survey paper they combined constraints from the DES itself with external SNe Ia, BAO, RSD, and Planck CMB with lensing: $\Omega_m=0.306$, $h=0.681$, $\Omega_\Lambda=0.694$, $\Omega_b=0.0486$, $\sigma_8=0.807$, $n_s=0.967$.} The summary of all the simulations used in this work is given in Tab.~\ref{tab:sims}. 

\begin{figure}
    \centering
    \includegraphics[width=0.8\linewidth]{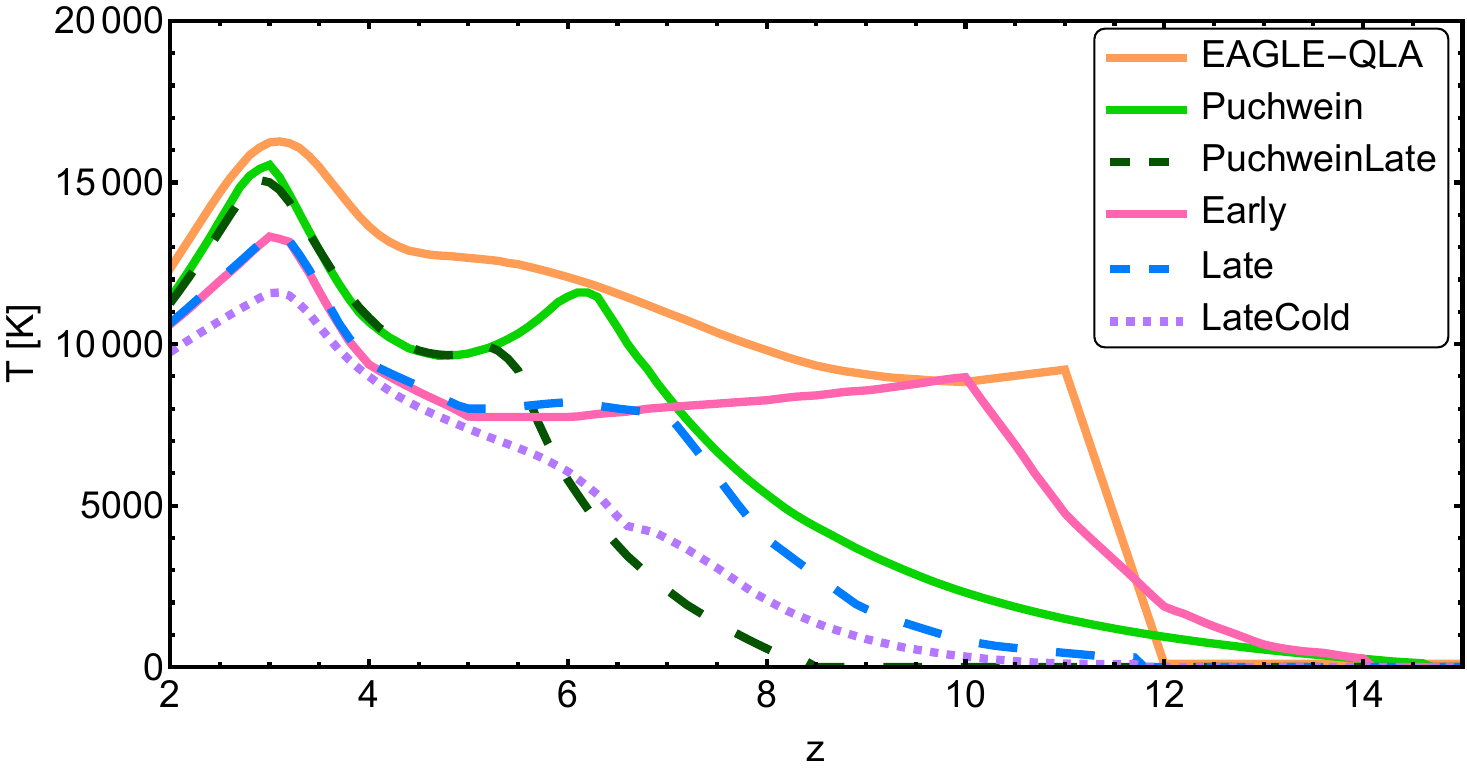}
    \caption{Temperature at the average density as a function of redshift for six different ultraviolet backgrounds analyzed in this paper. The hottest one (EAGLE-QLA -- solid orange line) is from~\cite{HM01}. A solid green line corresponds to the Puchwein et al.~\cite{puchwein2019consistent}, while the dark green dashed line is a modified Late Puchwein UVB (see Sec.~\ref{ssec:results_observations} for details). The final three (Early, Late, LateCold -- solid pink, dashed blue, and dotted violet lines) are from~\cite{onorbe2017self}.}
    \label{fig:UVB}
\end{figure}

The thermal histories of the simulations were varied using a simplified model with different redshift-dependent uniform ultraviolet backgrounds (UVBs). The modularity and open-source quality of the \textsc{swift} code allowed us to easily implement a routine that, self-consistently solves the heating/cooling rates equations from~\cite{katz1995cosmological}, similarly to~\cite{springel2003cosmological}. Fig.~\ref{fig:UVB} shows the temperature evolution at average density for each of the thermal histories used in this work.\footnote{This set of thermal histories was chosen as a broad range of thermal histories widely used in literature.}

Since the free streaming effect of WDM occurs prior to the initial redshift of our simulations, the imprints of this effect will be accounted for through modifications of the DM initial conditions (ICs). The ICs were computed using the open-source code \textsc{MonofonIC}~\cite{hahn2021higher} to redshift $z=30$. The effects of WDM were added to our ICs using the same approach as~\cite{viel2005constraining}, where the impact on the initial power spectrum depends solely on the mass of the dark matter particle. This was calculated using the Boltzmann solver \textsc{class}~\cite{blas2011cosmic} code which is already integrated into the \textsc{MonofonIC} code.

The resolution of the simulations naturally affects the size of the structures formed. Most of the simulations utilized, feature a box size of $\qty{25}{cMpc}$. Although this volume limits the formation of large structures, it does not have a significant impact on the flux power spectrum around the cut-off scales. By restricting the simulation volume, we were able to explore different resolutions with a broader range of number of particles. We found that a resolution of $2 \cdot 1024^3$ particles (equal parts dark matter and gas particles)\footnote{This resolution with box size of $\qty{25}{cMpc}$ corresponds to $\qty{9.1e4}{M_\odot}$ gas particle mass in the simulation.} provides a suitable balance, ensuring the necessary convergence at smaller and larger scales (see~\cite{bolton2009resolving,lukic2015lyman}), while retaining reasonable computational time. For further details about the resolution analysis, we refer the reader to the App.~\ref{app:resolution}. %For comparison with observations, however, this effect is negligible.

\begin{table}
    \centering
    \begin{tabular}{ccccc} \hline 
         Name&  $L_\mathrm{box}$ [cMpc]&  N&  UVB&  DM\\ 
         \hline
         EAGLE-QLA2048&  $25$&  $2048^3$&  HM01 \cite{HM01}&  CDM\\
         EAGLE-QLA1024&  $25$&  $1024^3$&  HM01 \cite{HM01}&  CDM\\
         EAGLE-QLA0752&  $25$&  $752^3$&  HM01 \cite{HM01}&  CDM\\
         EAGLE-QLA0512&  $25$&  $512^3$&  HM01 \cite{HM01}&  CDM\\
         EAGLE-QLA0376&  $25$&  $376^3$&  HM01 \cite{HM01}&  CDM\\
         \hline
         Puchwein1024&  $25$&  $1024^3$&  Puchwein et al. \cite{puchwein2019consistent}&  CDM\\
         Puchwein0752&  $25$&  $752^3$&  Puchwein et al. \cite{puchwein2019consistent}&  CDM\\
         Puchwein0512&  $25$&  $512^3$&  Puchwein et al. \cite{puchwein2019consistent}&  CDM\\
         Puchwein0376&  $25$&  $376^3$&  Puchwein et al. \cite{puchwein2019consistent}&  CDM\\
         \hline
         WDM-2.5keV-Puchwein&  $25$&  $1024^3$&  Puchwein et al. \cite{puchwein2019consistent}& $m=\qty{2.5}{keV}$\\
         WDM-3keV-PuchweinLate&  $25$&  $1024^3$&  PuchweinLate& $m=\qty{3}{keV}$\\
         \hline
         \hline
         Early&  $20/h$&  $1024^3$&  Early \cite{onorbe2017self}&  CDM\\
         Late&  $20/h$&  $1024^3$&  Late \cite{onorbe2017self}&  CDM\\         
         LateCold&  $20/h$&  $1024^3$&  LateCold \cite{onorbe2017self}&  CDM\\
         WDM-1keV& $20/h$& $1024^3$& None& $m=\qty{1}{keV}$\\
         WDM-2keV& $20/h$& $1024^3$& None& $m=\qty{2}{keV}$\\
         WDM-2p5keV& $20/h$& $1024^3$& None& $m=\qty{2.5}{keV}$\\
         WDM-3keV& $20/h$& $1024^3$& None& $m=\qty{3}{keV}$\\
         WDM-2keV-LateCold& $20/h$& $1024^3$& LateCold \cite{onorbe2017self}& $m=\qty{2}{keV}$\\ 
         \hline
    \end{tabular}
    \caption{Hydrodynamical simulations considered in this work, together with corresponding parameters. Columns contain from left to right: simulation identifier, comoving box size $L_\mathrm{box}$, number of gas particles $N$ (also equal to the number of dark matter particles), ultraviolet background used for the simulation, dark matter model in the simulation -- either CDM or the thermal relic model~\cite{viel2005constraining} with the corresponding particle mass.
    \newline
    All simulations are N-body SPH simulations that assume a uniform UVB. Simulations listed above the double line we run ourselves using \textsc{swift}~\cite{SWIFT}. EAGLE-QLA simulations were performed using default EAGLE-QLA flag (quick Lyman-alpha mode), with cooling from~\cite{wiersma2009effect} with UVB from~\cite{HM01}. Puchwein simulations are with Puchwein et al. UVB~\cite{puchwein2019consistent}. In WDM-3keV-PuchweinLate simulation, we used PuchweinLate UVB described below in Sec.~\ref{ssec:results_observations}. The
    initial conditions with different resolutions with warm and cold dark matter were prepared using \textsc{class}~\cite{blas2011cosmic} and \textsc{MonofonIC}~\cite{hahn2021higher}.
    The eight simulations listed below the double line are taken from the Garzilli et al. papers~\cite{garzilli2019lyman,garzilli2021constrain}, and were done using a modified \textsc{gadget-2}~\cite{springel2005cosmological}.}
    \label{tab:sims}
\end{table}

%%%%%%%%%%%%%%%%%%%%%%%%%%%%%%%%%%%%%%%%%%%%%%%%%%
\subsection{Analysis of the simulations}\label{ssec:simulations_analysis}
%%%%%%%%%%%%%%%%%%%%%%%%%%%%%%%%%%%%%%%%%%%%%%%%%%
For the analysis of the simulation data we used the open-source Python package \textsc{specwizard} ~\cite{spwz}. The code extracts random lines of sight (LOS) from the cosmological simulated volume,  computes an SPH interpolation to project the physical properties of the particles into pixels of a given size (usually fixed step in velocity space), and calculates the optical depth of the specific sightline for the desired set of ion transitions. Additionally, it includes post-processing tools such as the calculation of the flux power spectrum, flux probability density function, contaminant lines, and instrument noise. For more details, we refer the reader to ~\cite{spwz}.  

\begin{figure}
    \centering
    \includegraphics[width=1\linewidth]{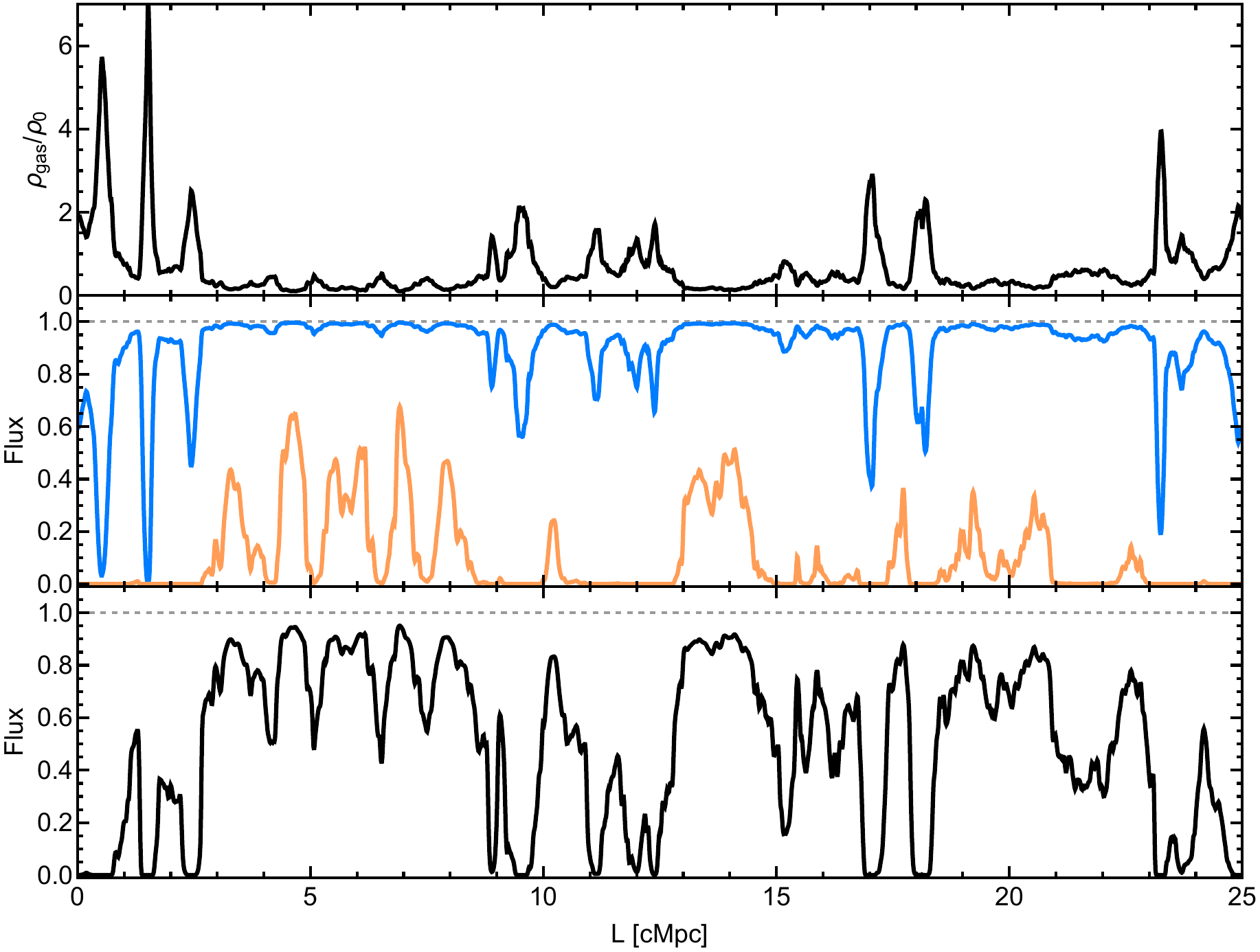}
    \caption{Example of a line of sight from EAGLE-QLA2048 simulation, $z=4.0$. The first panel shows normalized hydrogen density along a line of sight. The second panel is the real-space flux for the same density with different optical depth normalizations. The blue line is normalized so $\aver{F}=0.9$ and orange so $\aver{F}=0.1$. Note that the blue line significantly varies only in the regions of the highest density, while the yellow one is sensitive only to the lowest density regions. The third panel shows our default choice of flux normalization $\aver{F}=0.5$.}
    \label{fig:Freal}
\end{figure}

For each redshift snapshot, we used \textsc{specwizard} to calculate the neutral hydrogen optical depth of 3000 randomly positioned LOS.\footnote{In the case of Garzilli et al. simulations we had access only to 1000 LOS.} \textsc{specwizard} lets the user generate spectra with or without certain physical effects including the Doppler shifts of the lines due to peculiar velocities of the particles and/or thermal broadening of the optical depth. For most of our analysis, we removed these effects, so we were able to do a one-to-one mapping between the flux ($e^{-\tau}$) and the neutral hydrogen density to focus only on the pressure and WDM effects. This approach is equivalent to the `real-space flux' of~\cite{kulkarni2015characterizing}. We used the post-processing tool to rescale the optical depth and fix the desired average flux $\aver{F}$ of the computed sightlines. As we can observe in Fig.~\ref{fig:Freal}, different flux normalizations effectively probe regions of varying densities. For instance, when setting $\aver{F}=0.9$ most of the structures have negligible optical depth and $F\approx1$, and only the densest structure are not transparent and impact the FPS. Conversely, $\aver{F}=0.1$ targets voids with significantly lower-than-average densities, while other structures  saturate flux $F \approx 0$. We found that keeping $\aver{F}=0.5$ provides a good balance, primarily probing regions of average density. While limiting our analysis to $\aver{F}=0.5$ prevents direct comparison with observations, we will address this issue using an additional ``benchmark'' simulation in Sec.~\ref{ssec:results_observations}.

Lastly, we calculate the FPS for each line of sight and average them over each wavenumber. We fit the averaged FPS with our model Eq.~\eqref{eq:full_fit} to get the cut-off scale $k_\mathrm{cut}$ as a function of redshift. Before fitting, it is necessary to determine the range of scales to include in the fit, $k\leq k_\mathrm{max}$. We found that restricting fit up to cut-off scale ($k_\mathrm{max}=k_\mathrm{cut}$) gives consistent results across various simulations. More details on how we simultaneously fit FPS to find $k_\mathrm{cut}$ and restrict number of fitted points as $k \leq k_\mathrm{cut}$ can be found in the App.~\ref{app:fps_to_cutoff}.

\begin{figure}
    \centering
    \includegraphics[width=1\linewidth]{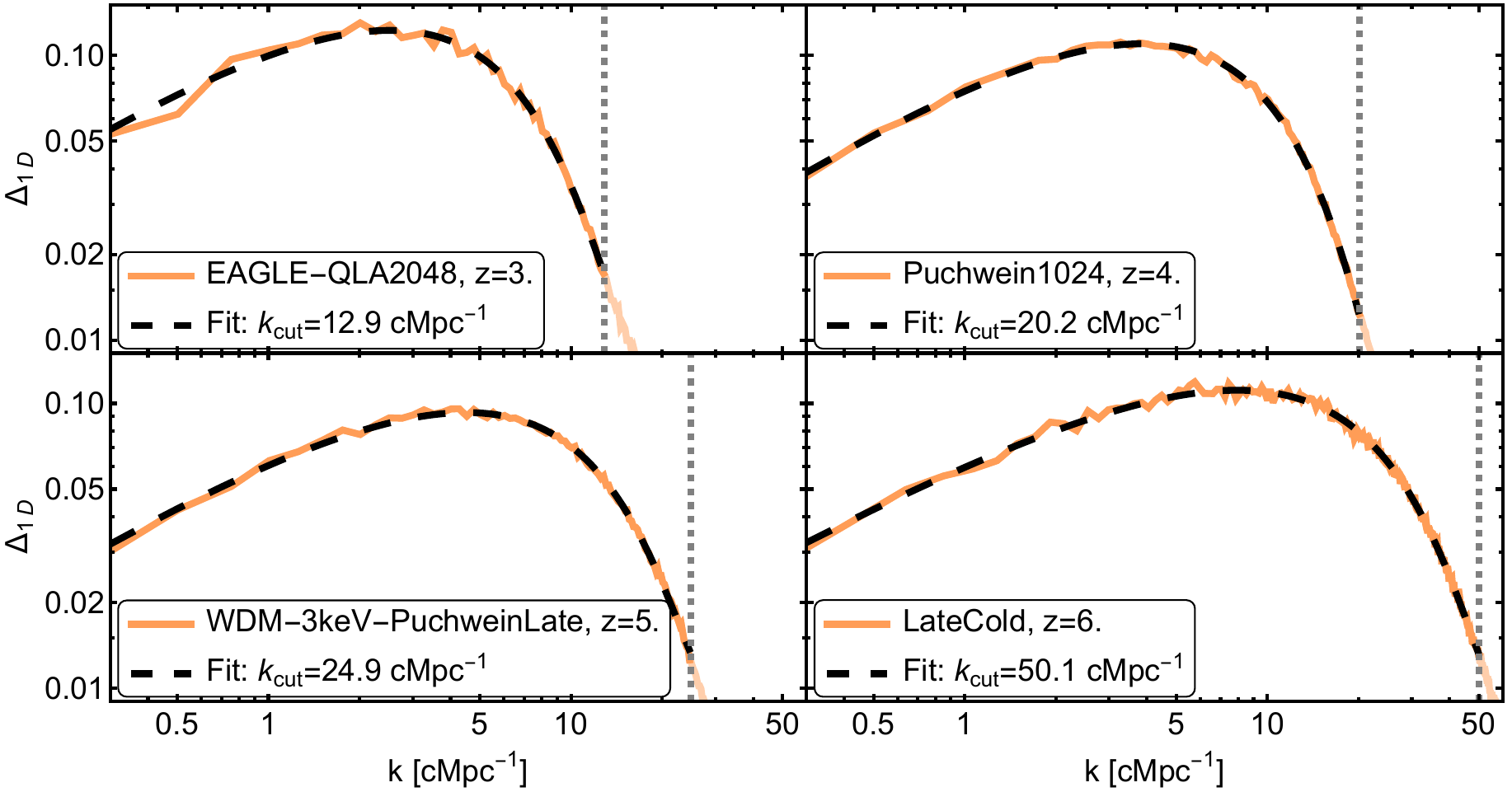}
    \caption{Examples of the cut-off in the real-space flux power spectrum. All four plots show 1D power spectra obtained from the simulations using the procedure described in Sec.~\ref{ssec:simulations_analysis} (orange line). Then we select points at scales $k<k_\mathrm{max}=k_\mathrm{cut}$ (darker orange part) and fit them using the function~\eqref{eq:full_fit} (black dashed line). For more details about fit see App.~\ref{app:fps_to_cutoff}. The vertical gray dashed line on each plot shows the corresponding cut-off scale $k_\mathrm{cut}$.}
    \label{fig:cut-off}
\end{figure}

On Fig.~\ref{fig:cut-off}, we shows examples of such fits. Different thermal histories with or without WDM across various redshifts with the filtering scale ranging from \mbox{$k_F\sim\qty{8}{cMpc^{-1}}$} to \mbox{$k_F\sim\qty{80}{cMpc^{-1}}$} are all fitted very well using the simple integral of the exponential cut-off~\eqref{eq:full_fit}.

%%%%%%%%%%%%%%%%%%%%%%%%%%%%%%%%%%%%%%%%%%%%%%%%%%

\section{Results}\label{sec:results}
%%%%%%%%%%%%%%%%%%%%%%%%%%%%%%%%%%%%%%%%%%%%%%%%%%

%%%%%%%%%%%%%%%%%%%%%%%%%%%%%%%%%%%%%%%%%%%%%%%%%%

\subsection{Relation between $\Delta_{\mathrm{3D}}$ and $\Delta_{\mathrm{1D}}$}\label{ssec:results_3D_to_1D}
%%%%%%%%%%%%%%%%%%%%%%%%%%%%%%%%%%%%%%%%%%%%%%%%%%

First, we test the accuracy of the translation between $\Delta_{\mathrm{3D}}$ and $\Delta_{\mathrm{1D}}$ in our ansatz~\eqref{eq:full_fit}. Since $\Delta_{\mathrm{3D}}$ lacks a cut-off in the nonlinear regime $z\lesssim15$~\cite{kulkarni2015characterizing}, such test is only possible at earlier redshifts in a simulation with already strong small-scale suppression. This can be achieved either by simulating an artificially early reionization, causing strong pressure smoothing, or by using a WDM simulation, as WDM effects are imprinted before nonlinear collapse. While~\cite{kulkarni2015characterizing} explored the first approach, we adopted the second, running a WDM simulation ($m = \qty{3}{keV}$) without UVB.

\begin{figure}
    \centering
    \includegraphics[width=1\linewidth]{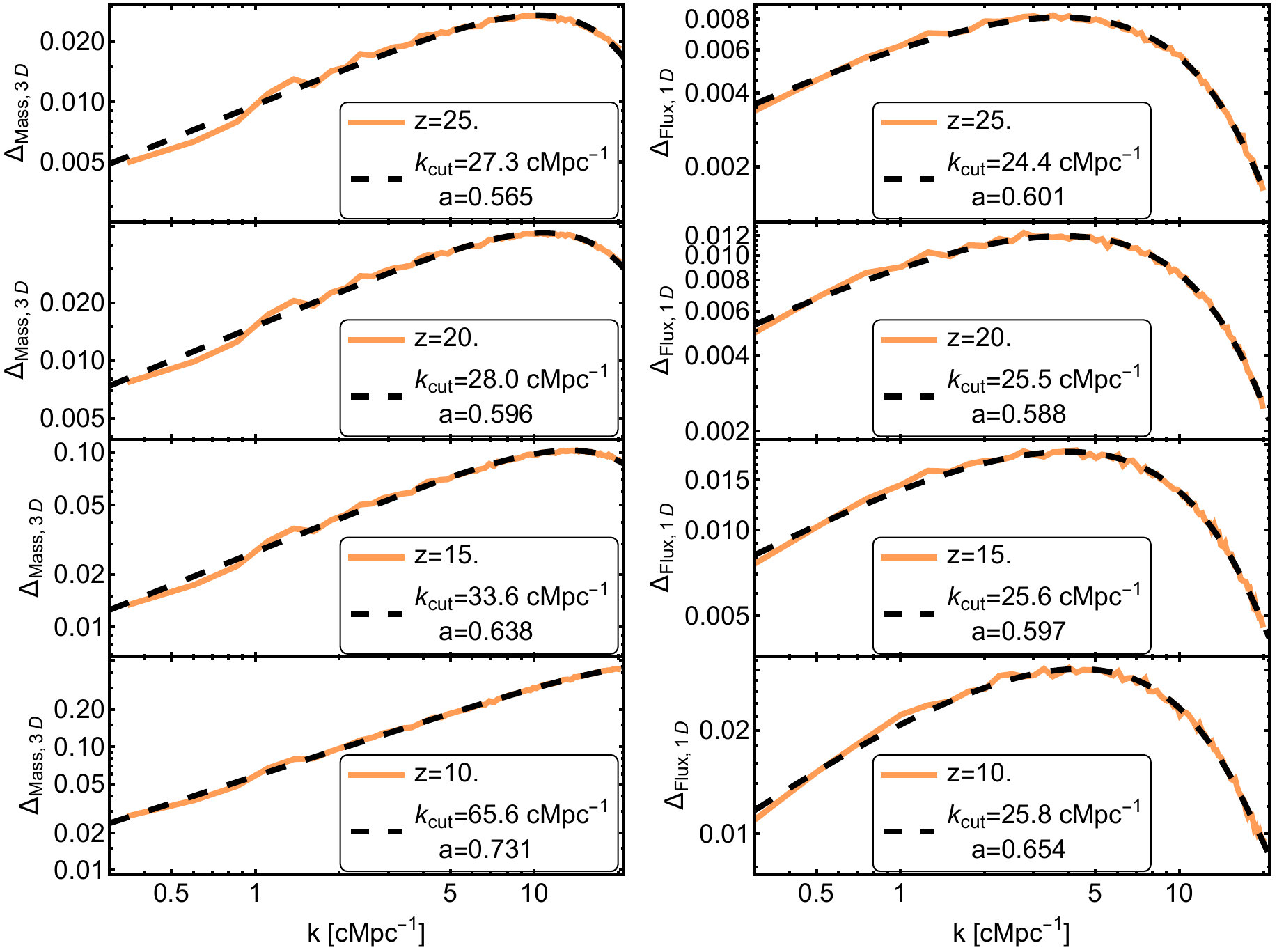}
    \caption{Comparison between 3D power spectrum (left side) and real-space flux power spectrum (right side). The model of WDM with $m=\qty{3}{keV}$ with no UVB was chosen. The first row corresponds to redshift $z=25$, second to $z=20$, third to $z=15$, and fourth to $z=10$. Orange lines are calculated from snapshots of simulations, black dashed lines are our fits. 3D matter power spectrum is calculated using \textsc{SWIFTSimIO}~\cite{borrow2020} and fitted using relation~\eqref{eq:full_exponent}. The 1D flux power spectrum is calculated as described in Sec.~\ref{ssec:simulations_analysis} and fitted by~\eqref{eq:full_fit}. For higher redshifts the power law slope $a$ and the cut-off scale $k_{\rm cut}$ are close between the 3D matter and 1D flux power spectra. For lower redshifts, nonlinear effects start to smooth out the cut-off in 3D matter PS (see~\cite{kulkarni2015characterizing}), while it remains almost unmodified in the 1D flux power spectrum and close to the theoretical estimate $k_\text{th}=\qty{27.1}{cMpc^{-1}}$, see Eq.~\eqref{eq:kWDM}. }
    \label{fig:3Dto1D}
\end{figure}

In Fig.~\ref{fig:3Dto1D}, we show the comparison between the 3D matter power spectrum and 1D flux power spectrum, both calculated from the simulation data and the fits from the theoretical model. At high redshifts ($z=25$ and $20$), we observe that both the location of the cut-off ($k_{\mathrm{cut}}$) and the power law slope ($a$) for the FPS and the 3D matter power spectrum are in good agreement. Moreover, the cut-off scale is in agreement with the theoretical estimate~\eqref{eq:kWDM} -- $k_\text{th}=\qty{27.1}{cMpc^{-1}}$. As we enter the regime where nonlinear effects are more prominent, the cut-off of the 3D matter power spectrum disappears while the 1D FPS cut-off stays approximately constant, as expected for the effect of WDM. This means that, for such redshift, we cannot study the cut-off from the matter PS, but we can using the flux PS; furthermore it gives us results consistent to ones predicted by linear theory for matter power spectrum.

%%%%%%%%%%%%%%%%%%%%%%%%%%%%%%%%%%%%%%%%%%%%%%%%%%
\subsection{Test of the pressure smoothing and WDM effects}\label{ssec:results_wdm_pressure}
%%%%%%%%%%%%%%%%%%%%%%%%%%%%%%%%%%%%%%%%%%%%%%%%%%

\begin{figure}
    \centering
    \includegraphics[width=\linewidth]{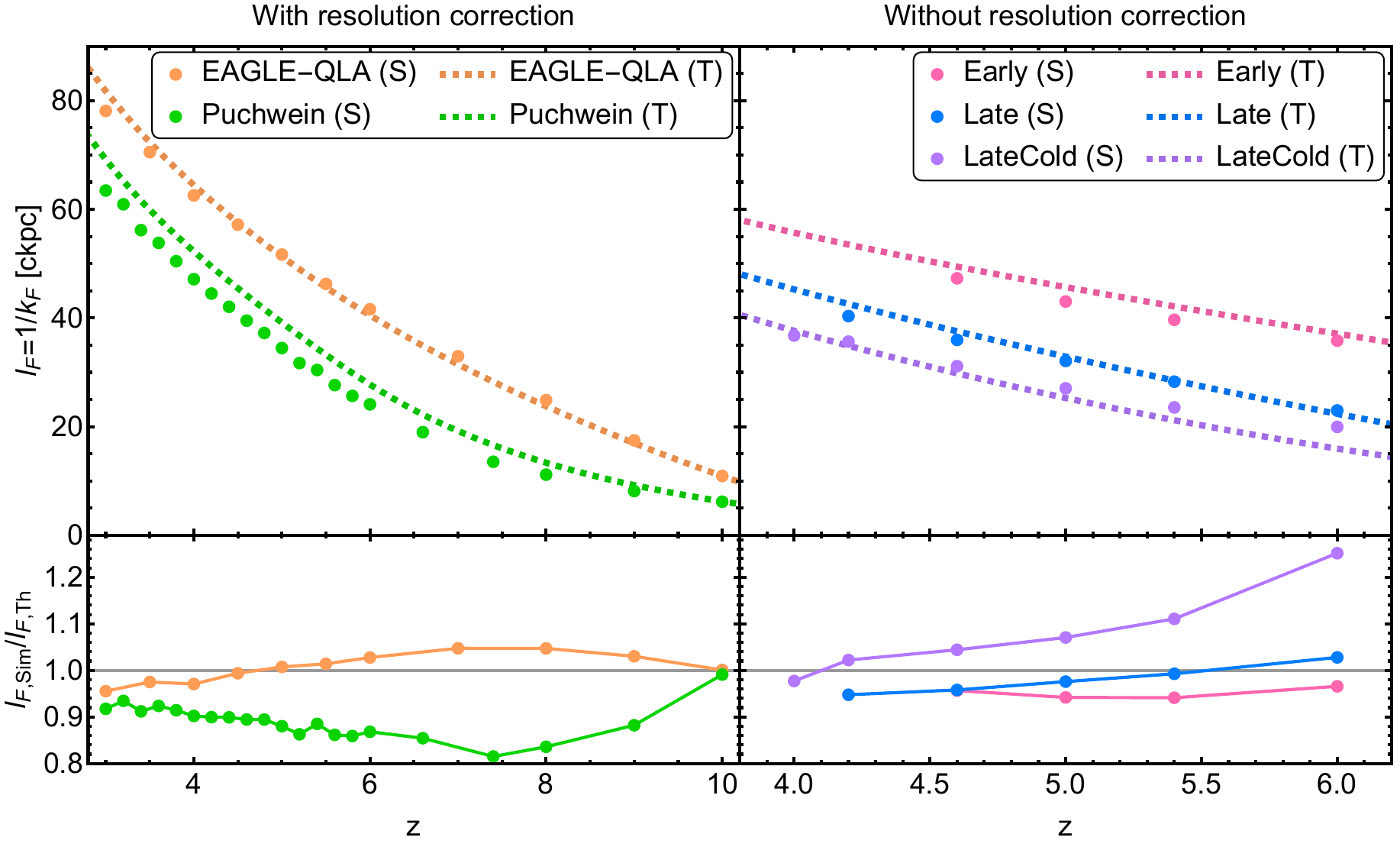}
    \caption{Filtering length as a function of redshift for different thermal histories in CDM model. The left panels show results from the simulations that we ran for this paper with resolution correction applied (see App.~\ref{app:resolution} for details). The right panels show results from the simulations by Garzilli et al.~\cite{garzilli2019lyman,garzilli2021constrain} without resolution correction. 
    The top panels display the redshift dependence of the filtering length for different simulations. Dashed lines are calculated using the \citetalias{gnedin1997probing} model (Eq.~\eqref{eq:kF}), while dots are our fits to the simulation data. The bottom panels depict the ratio between simulated and theoretical filtering lengths. It shows that most of the points are within $20\%$ of the \citetalias{gnedin1997probing} prediction.}
    \label{fig:pressure_results}
\end{figure}

In Fig.~\ref{fig:pressure_results}, we present a comparison between simulations and theoretical predictions for cases where the cut-off is only governed by pressure effect. The theoretical lines were generated by calculating the filtering length using the model of \citetalias{gnedin1997probing}, Eq.~\eqref{eq:kF}, with the temperature evaluated at the mean density, as illustrated in Fig.~\ref{fig:UVB}. The simulated filtering lengths were extracted from the simulations according to the procedure outlined in Sec.~\ref{ssec:simulations_analysis}. The lower panels indicate that most of simulated filtering scales fall within $20\%$ of the theoretical prediction across the redshift range $z=\{3-10\}$. The only exception is the LateCold scenario at $z=6$, likely due to resolution effects, as the cut-off scale for this point corresponds to a length of approximately $\sim \qty{20}{ckpc}$ which is below the resolution of this simulation. Two of our simulations (with EAGLE-QLA and Puchwein thermal history) were run in different resolutions and we are thus able to separate resolution effects from the pressure smoothing (left panel of Fig.~\ref{fig:pressure_results}). This significantly increased accuracy at earlier redshifts, where the cut-off scale is below resolution limit. For Garzilli et al.~\cite{garzilli2019lyman,garzilli2021constrain} simulations we had only one resolution run, therefore, we were not able to apply resolution correction (right panel of Fig.~\ref{fig:pressure_results}). For simulations with WDM we did not run different resolution setups, since the WDM cut-off, even at high redshifts, is significantly larger than the resolution we use. For an in-depth discussion of resolution effects and how we correct for them, see App.~\ref{app:resolution}.

\begin{figure}
    \centering
    \includegraphics[width=0.8\linewidth]{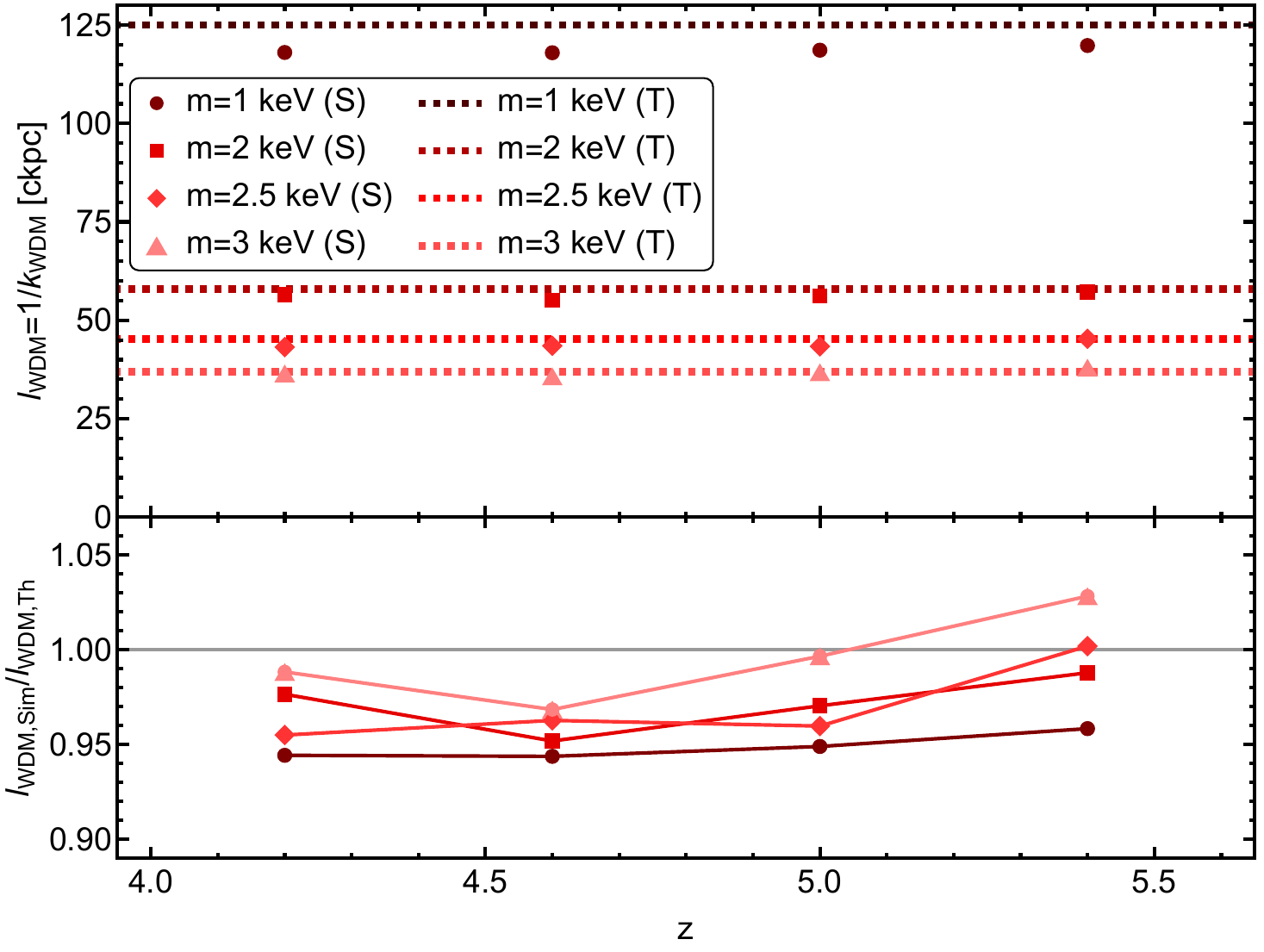}
    \caption{Filtering length as a function of redshift for different WDM models without UVB. The top panel shows a comparison of the WDM free streaming length from the simulations (dots) with the theoretical prediction (lines) from~\cite{viel2005constraining}. The bottom panel depicts the ratio between simulated and theoretical filtering length. Most of the points are within $\sim5\%$ of the theoretical model.}
    \label{fig:wdm_results}
\end{figure}

Analogously, in Fig.~\ref{fig:wdm_results} we explore the case in which the cut-off is only due to free streaming of WDM  particles with different masses. Most of the simulated cut-off lengths are within $5\%$ of theoretical values calculated by Eq.~\eqref{eq:kWDM} from~\cite{viel2005constraining}.

\begin{figure}
    \centering
    \includegraphics[width=\linewidth]{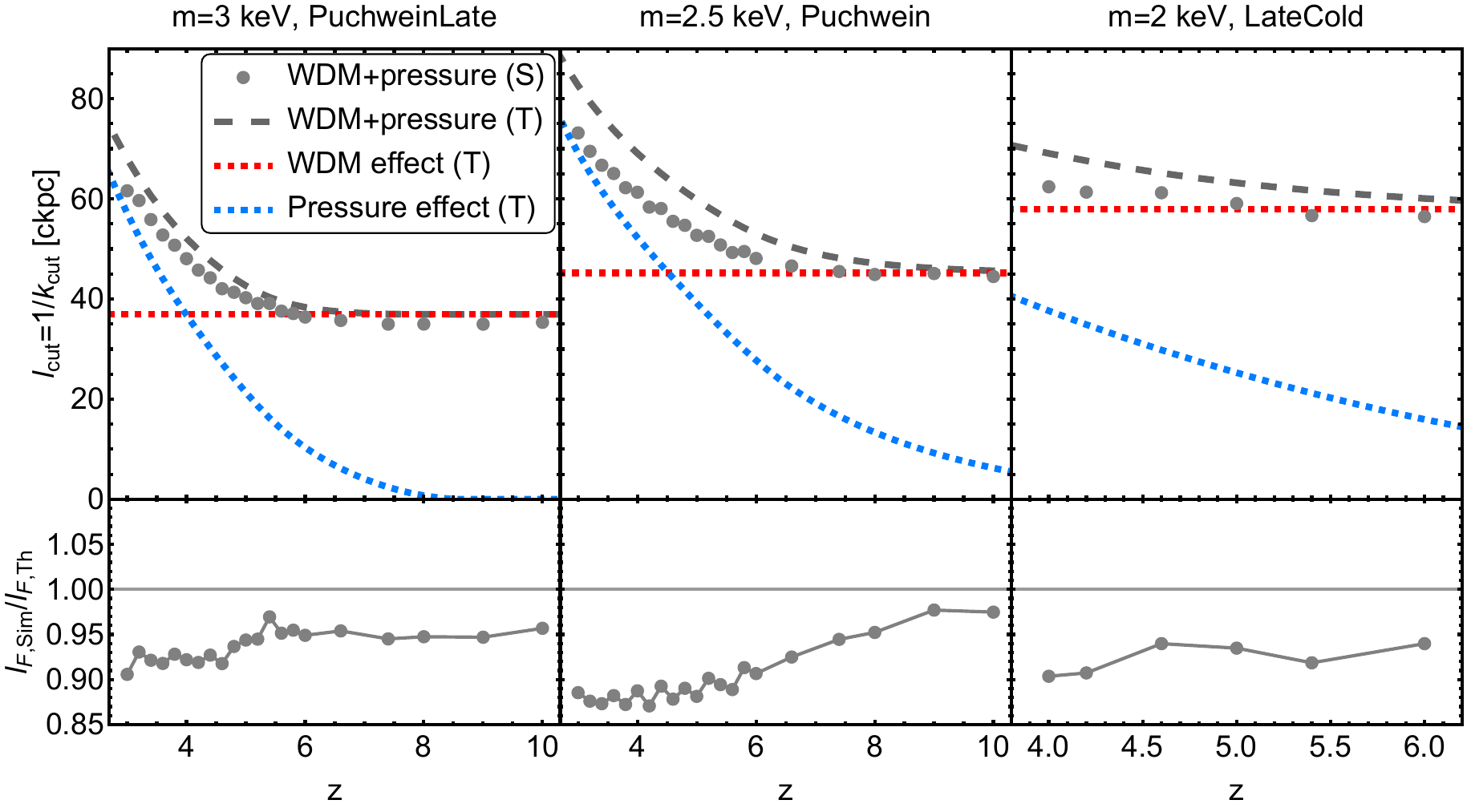}
    \caption{Comparison cut-off lengths in simulations with both WDM and pressure effects (gray dots) with the theoretical sum of two effects (gray dashed line): WDM effect (red dashed line) and pressure effect (blue dashed line) as described in Sec.~\ref{sec:theory}. The bottom panels show the ratio between simulation and theory.}
    \label{fig:adding_effects}
\end{figure}

Finally, in Fig.~\ref{fig:adding_effects}, we compare simulations that incorporate both WDM and pressure effects to theoretical predictions, as well as with our method of combining these effects, as described by Eq.~\eqref{eq:adding_effects}. The discrepancy between theory and simulations generally remains below $15\%$, likely resulting from minor differences between the theoretical and simulated pressure effects.

\subsection{Observations and predictive properties of the model}\label{ssec:results_observations}

Finally, we show how one can estimate the WDM particle mass in late reionization scenario that will correspond to the same cut-off as the pressure smoothing in early reionization. 

\begin{figure}
    \centering
    \includegraphics[width=\textwidth]{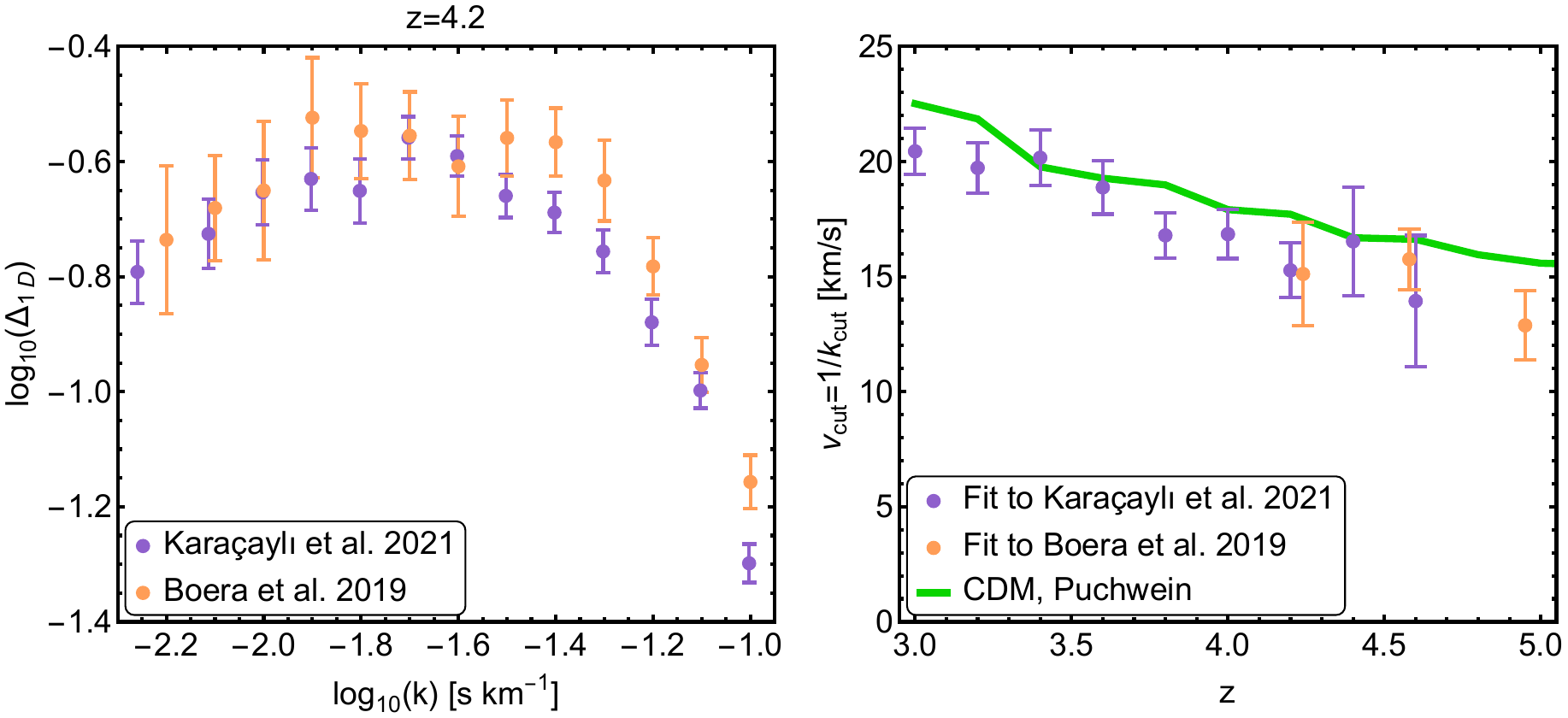}
    \caption{Left panel: comparison of flux power spectra from~\cite{karaccayli2022optimal} and~\cite{boera2019revealing} at redshift $z=4.2$. Right panel: cut-off length at different redshifts calculated from the observed FPS~\cite{karaccayli2022optimal, boera2019revealing} using fit~\eqref{eq:obs_fit} (dots) and simulated cut-off length from CDM with Puchwein UVB simulation (green line).}
    \label{fig:observations}
\end{figure}

We take the data from~\cite{karaccayli2022optimal,boera2019revealing}, where authors processed high-resolution quasar spectra and calculated the Lyman-apha flux power spectrum (see left panel of the Fig.~\ref{fig:observations} for an example of an observed FPS). Kara\c{c}ayl\i{} et al. 2021~\cite{karaccayli2022optimal} processed data from KODIAQ (HIRES spectrograph on the Keck I telescope), SQUAD (UVES spectrograph on the VLT telescope), XQ-100 (X-Shooter spectrograph on the VLT telescope) -- a total of 538 quasars at redshifts $z=2.0-4.6$. Boera et al. 2019~\cite{boera2019revealing} analyzed data from the HIRES spectrograph on the Keck I telescope and the UVES spectrograph on the VLT telescope -- a total of 15 quasars at redshifts $z=4.2-5.0$. Although this dataset is older and has fewer quasars, we include it because it probes higher redshifts.

It is crucial to note that in contrast with our previous analysis in which we used `real-space flux', observations include other effects like thermal broadening and peculiar velocity Doppler shifts. Thus the fit of the FPS cut-off had to be adjusted to Eq.~\eqref{eq:obs_fit}, for details we refer the reader to App.~\ref{app:observations_fit}. In the right panel of Fig.~\ref{fig:observations}, we show the cut-off scale $k_\mathrm{cut}$ redshift evolution by applying our theoretical fit on the data from~\cite{karaccayli2022optimal,boera2019revealing} (violet and orange dots). Note that for observational FPS, wavenumbers are traditionally in $\unit{s/km}$ units and we keep it this way for the cut-off scales here and below.

We start with our default simulation Puchwein1024. To compare it with observations we kept in \textsc{specwizard} all the physical effects on, and we used the mean flux fit from~\cite{boera2019revealing} to normalize the spectra from the simulation. We expect that the cut-off from this simulation will have a similar redshift dependence to observations since Puchwein et al.~\cite{puchwein2019consistent} constructed the UVB to reproduce observations. Indeed, this is the case, as the green solid line on the right panel in Fig.~\ref{fig:observations} shows, that matches closely the fits to the observational data.

From this analysis, we obtained the observational values of $k_\mathrm{cut}$ and the results of the CDM simulation. Now we can use our theoretical model to find the parameters for a WDM simulation that would reproduce the same cut-off, and hence, would also be in agreement with observations. To achieve this, we first designed a UVB background (labeled as PuchweinLate) that is identical to Puchwein et al.~\cite{puchwein2019consistent} for $z<5.3$ but with a shortened period of reionization that begins at $z=8.3$ instead of $z=15$. Because the heating of the universe starts later, this thermal history will have a smaller pressure effect. It is important to note that we are reducing the time of reionization but not the total amount of energy that is injected from the UVB. We can observe in Fig.~\ref{fig:UVB} that the line corresponding to PuchweinLate has a steeper slope during the reionization period but has the same temperature for $z<5.3$. 

\begin{figure}
    \centering
    \includegraphics[width=\linewidth]{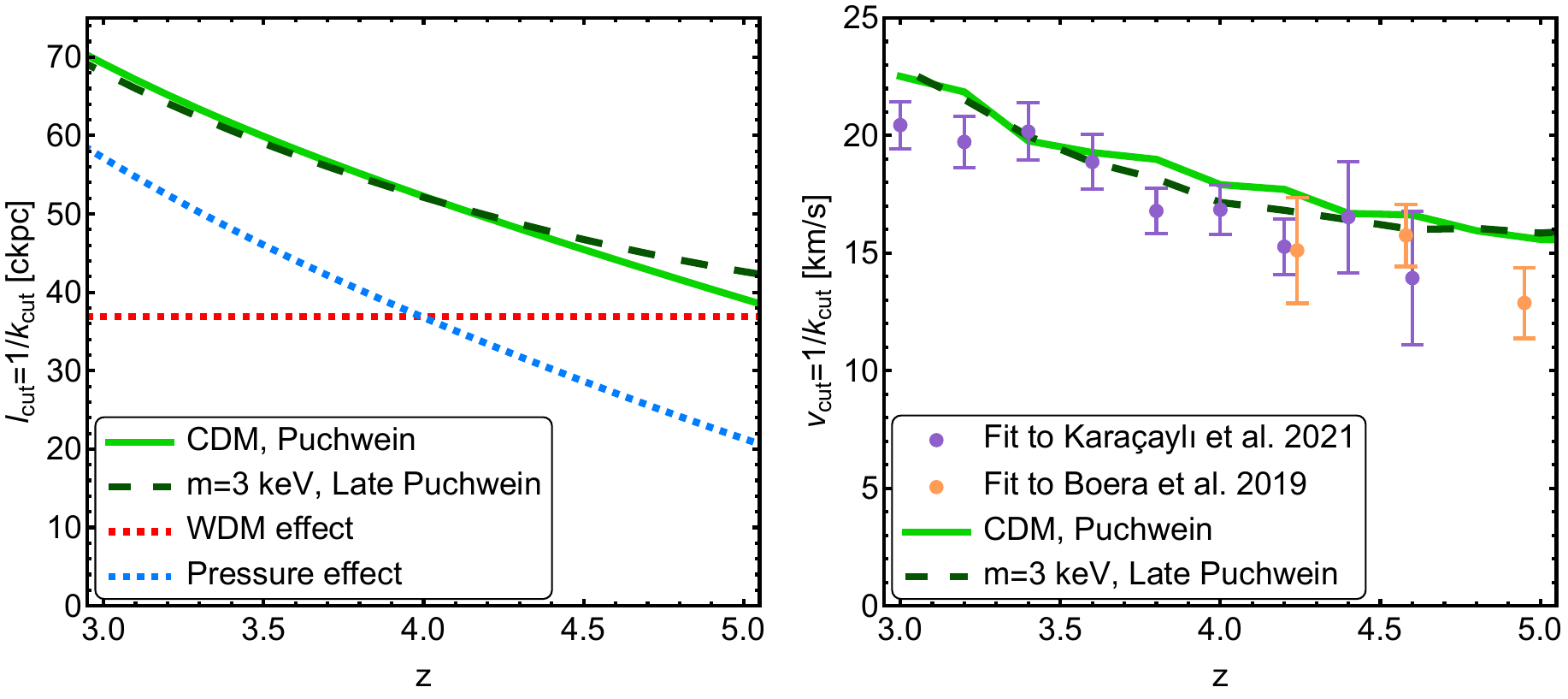}
    \caption{Cut-off lengths redshift evolution in CDM model with Puchwein thermal history (green solid line) and WDM $m=\qty{3}{keV}$ model with Late Puchwein thermal history (dark green dashed line). Left panel -- theoretical prediction for two models, where WDM $m=\qty{3}{keV}$ model with Late Puchwein is calculated as a sum~\eqref{eq:adding_effects} of pressure (blue dotted line) and WDM (red dotted line) effects. Right panel -- simulations results (lines) with all effects included compared with each other and observations (dots).}
    \label{fig:wdm_cdm}
\end{figure}

Now, using Eq.~\eqref{eq:kF} we can calculate the pressure smoothing effect for the original Puchwein and modified PuchweinLate thermal histories.\footnote{Note that for the pressure effect calculation, we need to know thermal history (meaning temperature dependence over time), while above we specified only the UVB. One can obtain the thermal history with high precision without running full simulation, by simply running a small box with uniform density and thus self-consistently solving the heating/cooling equations. We used this approach here too.} Then using the rule of adding effects -- Eq.~\eqref{eq:adding_effects}, we calculate the difference between pressure effects in both scenarios. Finally, with Eq.~\eqref{eq:kWDM} we calculate to what WDM particle mass this difference corresponds. In our calculations, the corresponding mass was $m=\qty{3}{keV}$. The left panel of Fig.~\ref{fig:wdm_cdm} schematically illustrates this computation. Note how close the solid green and dashed dark green lines are, even though they correspond to completely different scenarios: one with CDM and early reionization and another to WDM with later reionization. 

\begin{figure}
    \centering
    \includegraphics[width=\linewidth]{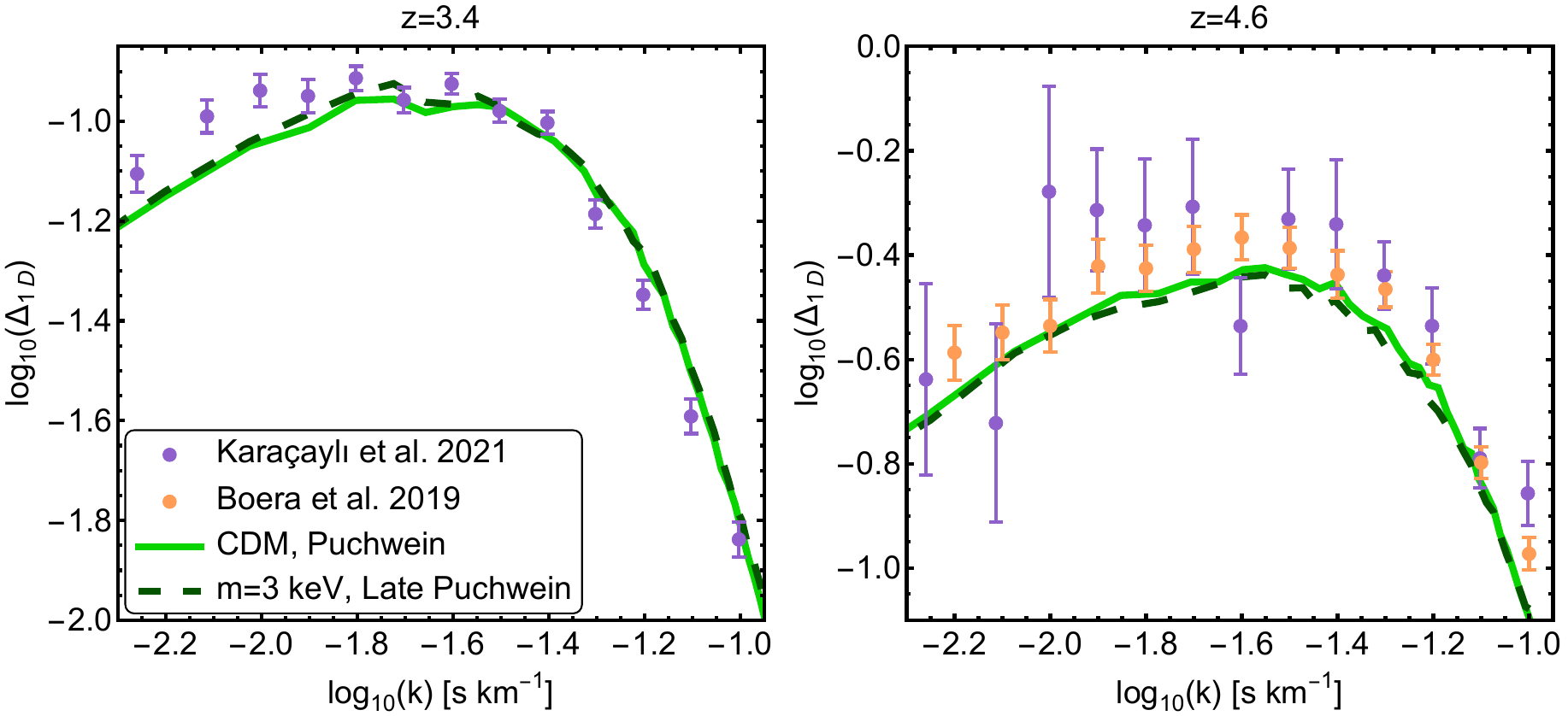}
    \caption{Comparison of observed FPS~\cite{karaccayli2022optimal} (violet dots) and~\cite{boera2019revealing} (orange dots) with FPS from CDM-Puchwein simulation (green line) and WDM-3keV-PuchweinLate (dark green dashed line) for redshifts $z=3.4$ (left panel) and $z=4.6$ (right panel).}
    \label{fig:fps_sim_data}
\end{figure}

We then run a corresponding simulation (labelled as WDM-3keV-PuchweinLate in Tab.~\ref{tab:sims}). The result from the simulation is shown in the right panel of Fig.~\ref{fig:wdm_cdm}. We observe in both cases that the difference between the CDM and the WDM simulations are within observational error bars. Importantly, these results do not allow us to clearly distinguish between a cut-off driven solely by pressure effects and one arising from a combined WDM+pressure simulation. Fig.~\ref{fig:fps_sim_data} directly compares FPS from observations and simulations and again show that the CDM and WDM simulations are indistinguishably close. Even though both simulations do not completely agree with the observations, the difference between two simulations is consistently below observational errors.

It should be emphasized that, even though this test involves greater complexity than tests using the real-space flux -- incorporating both thermal broadening and redshift-space distortions, our analytical recipe successfully predicted the similarities in the cut-off between simulations at different redshifts, with an error margin of less than 5\%. This success can be attributed to the fact that both thermal histories were modified only at redshifts $z>5$, maintaining similar levels of thermal line broadening at observed redshifts. Nevertheless, it is worth highlighting that our analytical approach provides predictive insights to explore various physical models without the need to run additional simulations.

%%%%%%%%%%%%%%%%%%%%%%%%%%%%%%%%%%%%%%%%%%%%%%%%%%
%%%%%%%%%%%%%%%%%%%%%%%%%%%%%%%%%%%%%%%%%%%%%%%%%%
    \section{Conclusions}
    \label{sec:conclusions}

In this study, we present a theoretical framework that aims to capture the combined effects of warm dark matter (WDM) free-streaming and pressure-induced filtering on the suppression of the small-scale 3D matter power spectrum. While this framework does not provide a first-principles derivation of the relation between the observed flux power spectrum of neutral hydrogen and the underlying matter distribution, it motivates a simple phenomenological ansatz (Eq. \ref{eq:full_fit}) for this connection. To test the validity of this ansatz, we perform and analyzed a suite of hydrodynamical simulations that explore a range of physical parameters affecting the shape and cutoff of the flux power spectrum.

As shown in Fig. \ref{fig:pressure_results}, when focusing solely on the pressure smoothing effect, the simulations aligned with our theoretical predictions within a 20\% margin.  In Fig. \ref{fig:wdm_results} our model demonstrated an accuracy of within 5\%, this for the effects
driven exclusively by warm dark matter. When accounting for the combined contributions of both free-streaming and pressure effects, the discrepancy between simulation results and Eq. \ref{eq:full_fit} was consistently below 15\%.

In Sec. \ref{ssec:results_observations}, we applied our theoretical framework to the flux power spectrum observational data from~\cite{karaccayli2022optimal,boera2019revealing} and as presented in Fig. \ref{fig:observations} we estimated their corresponding cut-off scales across different redshifts. We ran a cold dark matter (CDM) simulation with the thermal history from~\cite{puchwein2019consistent}, and post-processed the data to match the average flux fit from~\cite{boera2019revealing}. This CDM simulation was consistent with the observed data. Subsequently, as can be observed in Fig. \ref{fig:wdm_cdm}, we employed our theoretical model to design a thermal history and a WDM particle mass that could reproduce the flux power spectrum of the CDM simulation. Remarkably, as seen in Fig. \ref{fig:fps_sim_data}, the resulting flux power spectra were nearly indistinguishable, differing by less than 5\% at the cut-off scale.

The success of our phenomenological approach highlights several important open questions. Notably, there is no rigorous theoretical justification for why this method works. While a strict mathematical relationship exists between the 3D and 1D power spectra of the same quantity (see Eq.~\eqref{eq:1D-3D-connection}), our approach relies on a linearized theory of the matter power spectrum, which does not accurately describe the true matter power spectrum at low redshifts. Furthermore, we employ this approximation to predict the behavior of the 1D flux power spectrum, which is related to the matter distribution through a nonlinear transformation. The fact that this approach yields accurate results across a broad parameter space suggests the presence of an underlying, nontrivial connection that remains unexplored. Investigating this relationship further and modeling other physical effects present in the Ly-alpha forest are key objectives of our future research.

\section*{Acknowledgments}
Ivan Ridkokasha acknowledges financial support in the form of a scholarship from the Den Adel Fund. 
This work used the DiRAC@Durham facility managed by the Institute for Computational Cosmology on behalf of the STFC DiRAC HPC Facility (www.dirac.ac.uk). The equipment was funded by BEIS capital funding via STFC capital grants ST/P002293/1, ST/R002371/1 and ST/S002502/1, Durham University and STFC operations grant ST/R000832/1. DiRAC is part of the National e-Infrastructure. This work used the Dutch national e-infrastructure with the support of the SURF Cooperative using grant no. EINF-7355. 

%%%%%%%%%%%%%%%%%%%%%%%%%%%%%%%%%%%%%%%%%%%%%%%%%%
%%%%%%%%%%%%%%%%%%%%%%%%%%%%%%%%%%%%%%%%%%%%%%%%%%
%%%%%%%%%%%%%%%%%%%%%%%%%%%%%%%%%%%%%%%%%%%%%%%%%%
%%%%%%%%%%%%%%%%%%%%%%%%%%%%%%%%%%%%%%%%%%%%%%%%%%
\appendix

%%%%%%%%%%%%%%%%%%%%%%%%%%%%%%%%%%%%%%%%%%%%%%%%%%
%%%%%%%%%%%%%%%%%%%%%%%%%%%%%%%%%%%%%%%%%%%%%%%%%%
\section{Warm dark matter effect}\label{app:wdm_effect}
\begin{figure}
    \centering
    \includegraphics[width=0.8\linewidth]{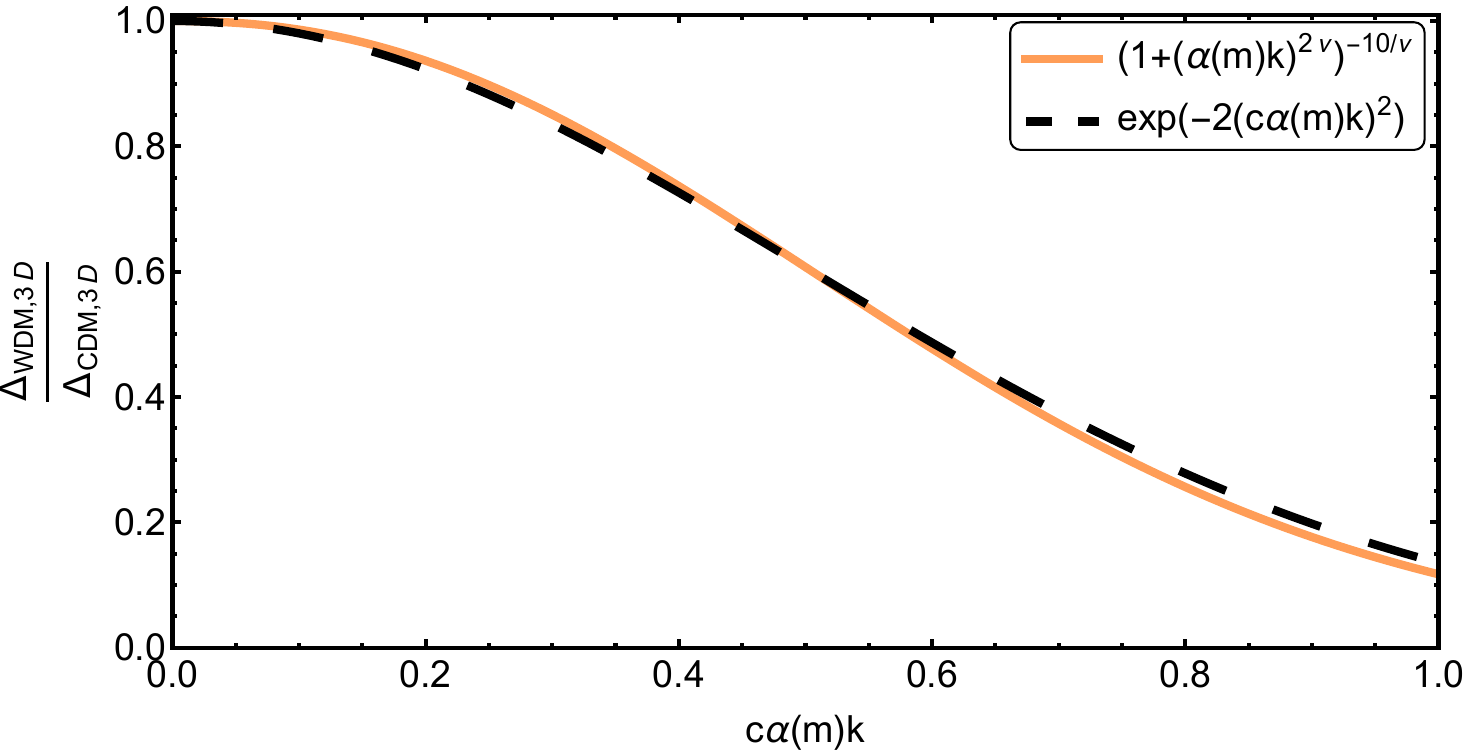}
    \caption{Comparison between parametrizations~\eqref{eq:vielwdm} and~\eqref{eq:exponent_wdm2} with $c=1.79$ of the WDM cut-off in thermal relic model. The maximal difference between two functions is 0.025 within the scales of interest: from $k=0$ to $k=\frac{1}{c\cdot\alpha(m)}=k_\mathrm{WDM}$.}
    \label{fig:wdm_coef}
\end{figure}

%%%%%%%%%%%%%%%%%%%%%%%%%%%%%%%%%%%%%%%%%%%%%%%%%%
%%%%%%%%%%%%%%%%%%%%%%%%%%%%%%%%%%%%%%%%%%%%%%%%%%
Throughout this paper, we use a model of thermal relic warm dark matter. Viel et al.~\cite{viel2005constraining} used a Boltzmann solver and ran it for different dark matter masses to estimate WDM effect on the 3D matter power spectrum. For thermal relics in linear regime, Viel et al. found that a good fit is
\begin{equation}\label{eq:vielwdm}
    \begin{aligned}
        \frac{\Delta^\mathrm{3D}_\mathrm{WDM}}{\Delta^\mathrm{3D}_\mathrm{CDM}}&=(1+(\alpha(m) k)^{2\nu})^{-10/\nu},\\
        \alpha(m)&=0.07 \left(\frac{m}{\qty{1}{keV}}\right)^{-1.11} \unit{Mpc},\\
        \nu&=1.12.
    \end{aligned}
\end{equation}
Note that the wavenumber enters the equation only in combination with $\alpha(m)$ as $\alpha(m)k$. To simplify the analysis we approximate the form of the cut-off with an exponent, but to retain the same dependence on the mass of the thermal relic we also enter wavenumber only in the same combination $\alpha(m)k$:
\begin{equation*}
    \frac{\Delta^\mathrm{3D}_\mathrm{WDM}}{\Delta^\mathrm{3D}_\mathrm{CDM}}=\exp{\left(-2(c\cdot\alpha(m) k)^2\right)}.
\end{equation*}
This translation has a free parameter $c$ that scales the whole cut-off. We found that the value $c=1.79$ produces a very similar form of the cut-off to~\eqref{eq:vielwdm} with a maximal deviation of 0.025 (see Fig.~\ref{fig:wdm_coef}). This corresponds to 
\begin{align}\label{eq:exponent_wdm2}
    \frac{\Delta^\mathrm{3D}_\mathrm{WDM}}{\Delta^\mathrm{3D}_\mathrm{CDM}}&=\exp{\left(-2\frac{k^2}{k_\mathrm{WDM}^2}\right)},\\
    k_\mathrm{WDM}&=\frac{1}{1.79\cdot\alpha(m)}=8.0\left(\frac{m}{\qty{1}{keV}}\right)^{1.11} \unit{Mpc^{-1}}.\label{eq:kWDM}
\end{align}
This effect is constant over redshifts of interest since dark matter free streaming happened long before the redshifts relevant to observations.

%%%%%%%%%%%%%%%%%%%%%%%%%%%%%%%%%%%%%%%%%%%%%%%%%%
%%%%%%%%%%%%%%%%%%%%%%%%%%%%%%%%%%%%%%%%%%%%%%%%%%
\section{Pressure smoothing effect}\label{app:pressure_effect}
%%%%%%%%%%%%%%%%%%%%%%%%%%%%%%%%%%%%%%%%%%%%%%%%%%
%%%%%%%%%%%%%%%%%%%%%%%%%%%%%%%%%%%%%%%%%%%%%%%%%%

In comoving coordinates, Gnedin \& Hui \citetalias{gnedin1997probing} showed that in linear regime in the limit $k\ll k_F$ the dimensionless 3D matter power spectra of baryons and DM relate as:
\begin{equation*}
    \frac{\Delta^\mathrm{3D}_\mathrm{baryons}(z,k)}{\Delta^\mathrm{3D}_\mathrm{DM}(z,k)}=1-2\frac{k^2}{k_F^2(z)},
\end{equation*}
with the filtering scale $k_F$ in matter dominated regime equal to
\begin{equation} \label{eq:kF}
    \frac{1}{k_F^2(z)}=\frac{3(1+z)^2}{k_{J}^2(z)}\int_{z}^{\infty}\frac{T(z')}{T(z)}\frac{1}{(1+z')^2}\left(1-\sqrt{\frac{1+z}{1+z'}}\right)\,dz'.
\end{equation}
$k_J(z)$ is the Jeans scale in the comoving coordinates:
\begin{equation*}
    k_J(z)=\frac{\sqrt{4\pi G\rho_0(z)}}{(1+z)c_s(z)},
\end{equation*}
where $z$ -- corresponding redshift, $\rho_0$ -- average matter density, $c_s=\sqrt{\frac{\gamma kT}{\mu}}$ -- speed of sound. We take mean molecular mass $\mu\approx0.61 m_\mathrm{H}$, which corresponds to a fully ionized hydrogen and singly ionized helium, which is mostly true for the redshifts of interest.

For our analysis, we need to extend \citetalias{gnedin1997probing} prediction from $k\ll k_F$ to the higher wavenumbers $k\sim k_F$. For this, we need a model of the form of the cut-off.

It is difficult to analytically derive the form of the cut-off in the matter power spectrum. However, numerical solutions~\cite{gnedin2003linear}, show that for 3D matter power spectrum the following exponent closely follows simulations at $k\lesssim k_F$:
\begin{equation*}
    \frac{\Delta^\mathrm{3D}_\mathrm{baryons}}{\Delta^\mathrm{3D}_\mathrm{DM}}=\exp{\left(-2\frac{k^2}{k_F^2}\right)}.
\end{equation*}

%%%%%%%%%%%%%%%%%%%%%%%%%%%%%%%%%%%%%%%%%%%%%%%%%%
%%%%%%%%%%%%%%%%%%%%%%%%%%%%%%%%%%%%%%%%%%%%%%%%%%
\section{From FPS to cut-off}\label{app:fps_to_cutoff}
%%%%%%%%%%%%%%%%%%%%%%%%%%%%%%%%%%%%%%%%%%%%%%%%%%
%%%%%%%%%%%%%%%%%%%%%%%%%%%%%%%%%%%%%%%%%%%%%%%%%%

To determine the cut-off scale $k_\mathrm{cut}$ for each simulation, we fit the real-space FPS obtained as described in Sec.~\ref{ssec:simulations_analysis} with a function~\eqref{eq:full_fit}.

Before fitting, it is necessary to determine the range of scales to include in the fit, $k\leq k_\mathrm{max}$ (also see Sec.~\ref{sec:theory} about $k_\mathrm{max}$). We found that $k_\mathrm{max}=k_\mathrm{cut}$ works consistently well for different redshifts and thermal histories. Since the values of $k_\mathrm{cut}$ and $k_\mathrm{max}$ are not known in advance, we define them algorithmically:

\begin{enumerate}
    \item We start from a power spectrum -- a set of points $\{(k_{j},\Delta_{j})\ |\ j=1,2,..., N\}$, where $N$ is the number of points, $\{k_j,\Delta_j\}$ -- wavenumber and dimensionless power spectrum per point. $k_{j}$ in this sequence is in the ascending order.
    \item We create a sequence of fits:
    \begin{enumerate}
        \item 0-th fit: $k_{\mathrm{max},0}=k_{N}$ -- we fit all points. Result of a fit we save as $k_{\mathrm{cut},0}$;
        \item 1-st: $k_{\mathrm{max},1}=k_{N-1}$ -- we fit all points except the last one. We get $k_{\mathrm{cut},1}$;
        \item i-th: $k_{\mathrm{max},i}=k_{N-i}$ -- we fit the first $N-i$ points. We get $k_{\mathrm{cut},i}$.
    \end{enumerate}

    \item Then, we find a fit $i_{0}$ from the sequence, that satisfies $k_{\mathrm{max}, i_{0}}\approx k_{\mathrm{cut},i_{0 }}$. More precisely -- smallest $i$, when $k_{\mathrm{max}, i}=k_{N-i}< k_{\mathrm{cut},i}$. Then the filtering scale is defined as $k_{\mathrm{cut}}=k_{\mathrm{cut},i_{0}}$.
\end{enumerate}

%%%%%%%%%%%%%%%%%%%%%%%%%%%%%%%%%%%%%%%%%%%%%%%%%%
%%%%%%%%%%%%%%%%%%%%%%%%%%%%%%%%%%%%%%%%%%%%%%%%%%
\section{Resolution effects}\label{app:resolution}
%%%%%%%%%%%%%%%%%%%%%%%%%%%%%%%%%%%%%%%%%%%%%%%%%%
%%%%%%%%%%%%%%%%%%%%%%%%%%%%%%%%%%%%%%%%%%%%%%%%%%
Most of the analyzed simulations have $N=1024^3$ gas particles and the same number of dark matter particles in a box size of $L_\mathrm{box}=\qty{25}{cMpc}$ or $\qty{20}{cMpc/h}$ (see Tab.~\ref{tab:sims}).
While this spatial resolution, $l_\mathrm{res}=L_\mathrm{box}/N_\mathrm{gas}^{1/3}\approx\qty{25}{ckpc}-\qty{20}{ckpc/h}$, is adequate for most Lyman-$\alpha$ analyses~\cite{bolton2009resolving,lukic2015lyman}, recent findings~\cite{doughty2023convergence} suggest that higher resolution may be required for the convergence of the pressure effect at higher redshifts $z\gtrsim5.5$. This is especially prominent in our CDM runs at high redshifts, because the cut-off length becomes as small as $\qty{10}{ckpc}$ (see Fig.~\ref{fig:pressure_results}), below the resolution of our simulations.

To explore the resolution effect on our results, we performed additional runs with different resolutions (see Tab.~\ref{tab:sims}) for simulations with cut-off scales comparable to resolution. Namely, CDM runs in $L_\mathrm{box}=\qty{25}{cMpc}$ box size with $N=376^3,\ 512^3,\ 752^3$ gas particles for Puchwein thermal history and EAGLE-QLA. Additionally, we performed one high-resolution run $N=2048^3$ gas particles with EAGLE-QLA.

\begin{figure}
    \centering
    \includegraphics[width=\linewidth]{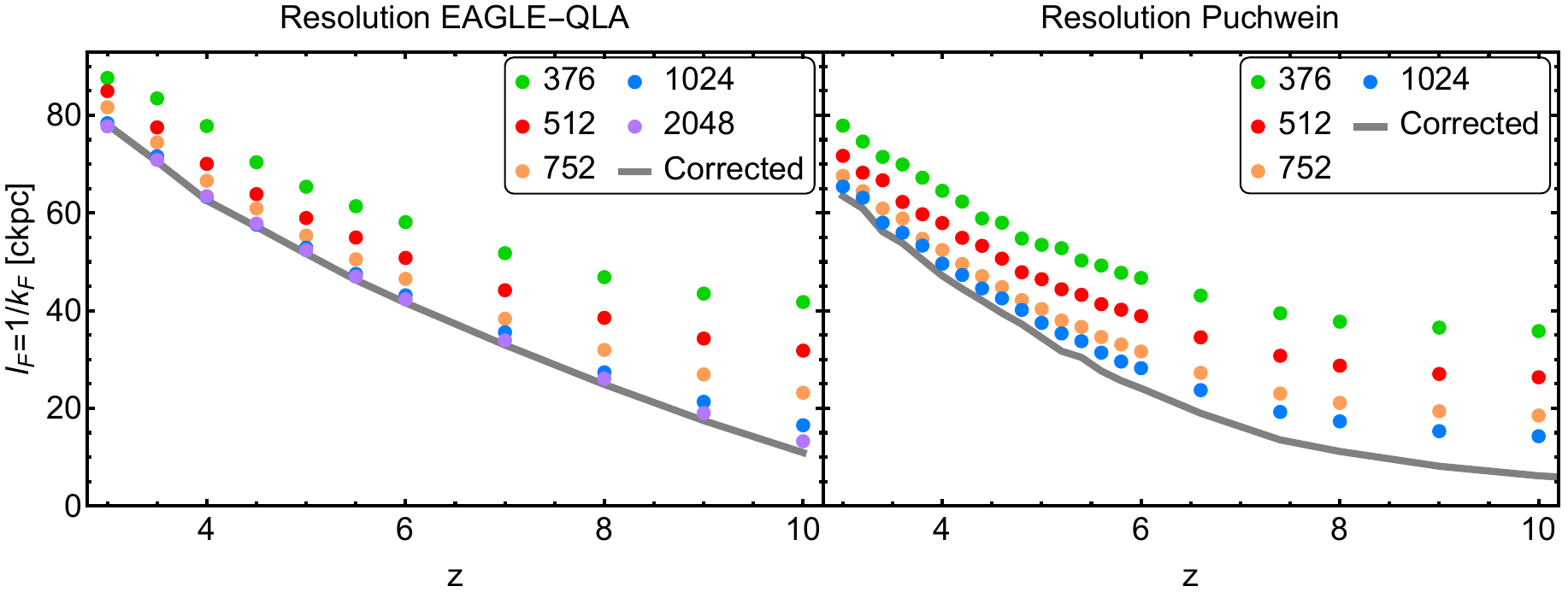}
    \caption{Comparison of EAGLE-QLA and Puchwein runs with different resolutions (dots). Solid gray line shows cut-off scale after resolution correction. Smaller resolutions have larger cut-off length, and difference become especially prominent at higher redshifts, as discussed in~\cite{doughty2023convergence}.}
    \label{fig:resolution}
\end{figure}

Using the same procedure as described in Sec.~\ref{ssec:simulations_analysis}, we compute the cut-off scale for each simulation at each redshift and plot it in Fig.~\ref{fig:resolution}. We see that lower resolutions result in higher cut-off lengths. The difference becomes especially prominent at high redshifts, when the pressure effect is relatively small, similar to the results of~\cite{doughty2023convergence}.

To understand this behavior, we can model the resolution effect as an additional exponential cut-off applied to the 3D power spectrum, where the cut-off length is proportional to the resolution:

\begin{align}
    \frac{\Delta^\mathrm{3D}_\mathrm{real}}{\Delta^\mathrm{3D}_\mathrm{ideal}}&=\exp{\left(-2\frac{k^2}{k_\mathrm{res}^2}\right)},\nonumber\\
    k_\mathrm{res}&=\frac{1}{\beta l_\mathrm{res}}=\frac{1}{\beta} \frac{N_\mathrm{gas}^{1/3}}{L_\mathrm{box}},\label{eq:kres}
\end{align}
where $\beta$ is an unknown dimensionless parameter. Then, after substituting~\eqref{eq:full_exponent} as $\Delta^\mathrm{3D}_\mathrm{ideal}$ we get
\begin{equation*}
    \frac{1}{k_\mathrm{sim}^2}=\frac{1}{k_\mathrm{cut}^2}+\frac{1}{k_\mathrm{res}^2},
\end{equation*}
where $k_\mathrm{sim}$ -- is the filtering scale we expect to obtain from fitting simulated FPS. Or rewriting it in cut-off lengths $l_\mathrm{cut}=1/k_\mathrm{cut}$ and substituting~\eqref{eq:kres}:
\begin{equation}\label{eq:adding_res}
    l_\mathrm{sim}^2=l_\mathrm{cut}^2+\beta^2\frac{L_\mathrm{box}^2}{N_\mathrm{gas}^{2/3}}.
\end{equation}

\begin{figure}
    \centering
    \includegraphics[width=0.8\linewidth]{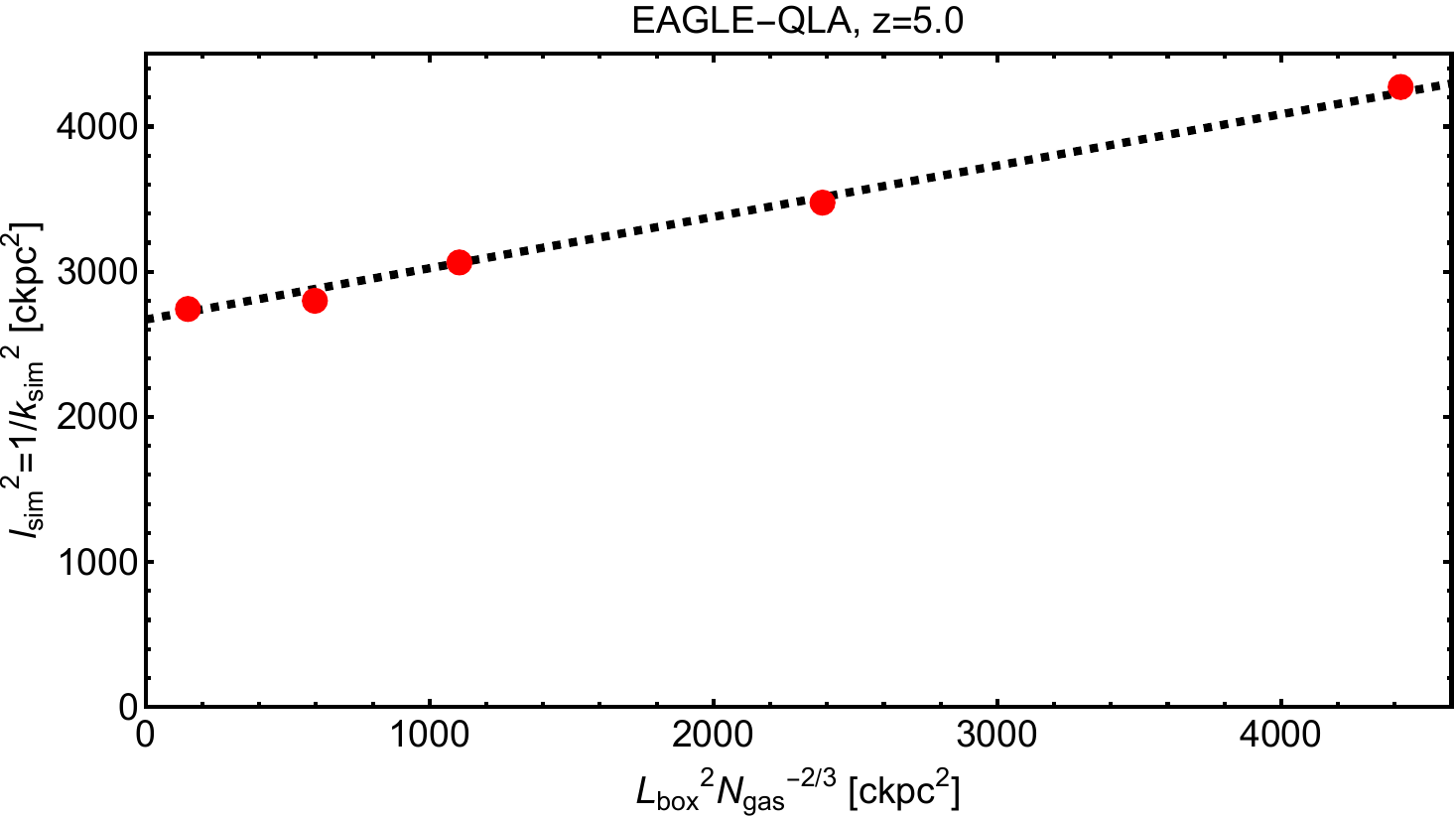}
    \caption{Example of a change of cut-off scale as a function of the resolution of the simulation, see Eq.~\eqref{eq:adding_res}. Red points show values of cut-off $l_\mathrm{sim}^2$ obtained from simulations with corresponding resolution. The dashed black line is a simple linear fit. We see that data perfectly lies on the line, meaning that our simple prescription of resolution effect works quite well. This fit allows us to estimate $k_\mathrm{cut}$ without the resolution effect by extrapolating $L_\mathrm{box}^2N_\mathrm{gas}^{-2/3}$ parameter to zero.}
    \label{fig:res_fit}
\end{figure}

In all the runs of the same thermal history we expect $l_\mathrm{cut}$ to be the same, and the only difference in observed cut-off $l_\mathrm{sim}$ will be because of the different resolution. Using a simple linear fit we can estimate both underlying ideal cut-off $l_\mathrm{cut}$ and unknown parameter $\beta$. An example of such fit is presented in the Fig.~\ref{fig:res_fit}. By repeating such fits at all redshift for both simulations we got the left panel of the Fig.~\ref{fig:pressure_results}.

%%%%%%%%%%%%%%%%%%%%%%%%%%%%%%%%%%%%%%%%%%%%%%%%%%
%%%%%%%%%%%%%%%%%%%%%%%%%%%%%%%%%%%%%%%%%%%%%%%%%%
\section{Fitting observational data}\label{app:observations_fit}
%%%%%%%%%%%%%%%%%%%%%%%%%%%%%%%%%%%%%%%%%%%%%%%%%%
%%%%%%%%%%%%%%%%%%%%%%%%%%%%%%%%%%%%%%%%%%%%%%%%%%

Compared to our analysis of pressure and WDM effects only, where we used `real-space flux', observational data includes also other physical effects like thermal broadening and Doppler shifts. Therefore, we need to adjust our previous ansatz~\eqref{eq:full_fit} and the whole fitting procedure from App.~\ref{app:fps_to_cutoff}. 

First, we conservatively limit the data from both~\cite{karaccayli2022optimal} and~\cite{boera2019revealing} to $k\leq\qty{0.1}{s/km}$ for all redshifts. Second, we found that the following equation gives an excellent fit for different scales and redshifts for both observational and simulated data: 
\begin{equation}\label{eq:obs_fit}
   \Delta_\mathrm{F}=Ak^a\frac{1}{\left(1+\frac{k^2}{k_\mathrm{cut}^2}\right)^2}.
\end{equation}
Correspondingly, this equation is our definition of cut-off scale $k_\mathrm{cut}$ for the observational data. To get the value and standard error of the cut-off scale we fit observed FPS at $k\leq\qty{0.1}{s/km}$ with an equation~\eqref{eq:obs_fit} using a standard least-squares method. To have a consistent estimate from simulations we also limit simulated data to values of $k\leq\qty{0.1}{s/km}$ for all redshifts and fit it with the same equation~\eqref{eq:obs_fit}.

\bibliographystyle{JHEP}
\bibliography{citations}

\end{document}